\documentclass[12pt,english,floatfix,nofootinbib,superscriptaddress,aps,prd,preprint]{revtex4}

\usepackage[utf8]{inputenc}
\usepackage{float}
\usepackage{array}
\usepackage{bbold}
\usepackage{lipsum}
\usepackage{dsfont}
\usepackage{graphicx}
\usepackage{amsmath,amsthm,amsfonts,amssymb}
\usepackage{graphicx}
\usepackage[english]{babel} 
\usepackage{color}
\usepackage{tensor}
\usepackage{esint}
\usepackage[dvips]{epsfig}
\usepackage[dvips]{graphicx}
\usepackage{float}
\usepackage{units}
\usepackage{textcomp}
\usepackage{mathrsfs}
\usepackage{amsmath}
\usepackage[makeroom]{cancel}
\usepackage{amssymb}
\usepackage{amsbsy}
\usepackage{amsfonts}
\usepackage{amssymb,mathrsfs,xcolor}
\usepackage{esint}
\usepackage{braket}
\usepackage{array}
\usepackage{graphicx}

\usepackage{wasysym}
\usepackage{multirow}
\usepackage{wrapfig}
\usepackage{subfig}

\usepackage{stmaryrd}
\usepackage{upgreek}

\makeatletter

\makeatletter\usepackage{babel}

\usepackage{hyperref}
\hypersetup{
    colorlinks,
    citecolor=blue,
    filecolor=green,
    linkcolor=purple,
    urlcolor=red,
}

\usepackage{slashed}

\newcommand{\ie}{\begin{equation}}
\newcommand{\fe}{\end{equation}}
\newcommand{\se}{\begin{eqnarray}}
\newcommand{\ff}{\end{eqnarray}}

\begin{document}

\title{Modified particle dynamics and thermodynamics in a traversable wormhole in bumblebee gravity}

%%%%%%%%%%%%%%%%%%%%%%%%%%%%%%%%%%%%%%%%%%%%%%%%%%%%%%%%%%%%%%%%%%%%%%
\author{A. A. Ara\'{u}jo Filho}
\email{dilto@fisica.ufc.br}

\affiliation{Departamento de Física, Universidade Federal da Paraíba, Caixa Postal 5008, 58051-970, João Pessoa, Paraíba,  Brazil.}

%%%%%%%%%%%%%%%%%%%%%%%%%%%%%%%%%%%%%%%%

\author{J. A. A. S. Reis}
\email{jalfieres@gmail.com}

\affiliation{Universidade Estadual do Sudoeste da Bahia (UESB), Departamento de Ciências Exatas e Naturais, Campus Juvino Oliveira, Itapetinga -- BA, 45700-00, Brazil.}

%%%%%%%%%%%%%%%%%%%%%%%%%%%%%%%%%%%%%%%%

\author{Ali \"Ovg\"un}
\email{ali.ovgun@emu.edu.tr}
\affiliation{Physics Department, Eastern Mediterranean
University, Famagusta, 99628 North Cyprus, via Mersin 10, Turkiye.}

%\author{L. Lisboa--Santos}
%\email{let\_lisboa@hotmail.com} 

%\affiliation{Universidade Federal do Cear\'a (UFC), Departamento de F\'isica,\\ Campus do Pici,
%Fortaleza - CE, C.P. 6030, 60455-760 - Brazil.}

%%%%%%%%%%%%%%%%%%%%%%%%%%%%%%%%%%%%%%%%%%%%%%%%%%%%%%%%%%%%%%%%%%%%%%%%%%%%%%%%%%%%%%%%%%%%%%%%%%%%%%%%%%%%%%%%%%%%%%%%%%%%%%%%%%%%%%%%%%%%%%%%%%%%%%%%%%%%%%%%%%%%%%%%%%%%%%%%%%%%%%%%%%%%%%%%%%%%%%%%%%%%%%%%%%%%%%%%%%%%%%%%%%%%%%%%%%%%%%%%%%%%%%%%%%%%%%%%%%%%%%%%%%%%%%%%%%%%%%%%%%%%%%%%%%%%%%%%%%%%%%%%%%%%

\date{\today}

\begin{abstract}

In this work, we analyze various phenomena influenced by the gravitational field in a bumblebee gravity solution, with a particular emphasis on a traversable wormhole for massless particle modes. Specifically, we calculate the index of refraction, group velocity, time delay, modified distances, and interparticle potential, demonstrating the possibility of photon--photon interactions due to the wormhole geometry. For the latter aspect, we also extend the analysis to massive particle modes, resulting in a ``combination'' of modified Yukawa-- and Couloumb--like potentials. These calculations are shown to be dependent on the wormhole's parameters, particularly the wormhole throat. In addition to these analyses, the Hawking temperature is derived using the trapping horizon method, yielding negative values. Furthermore, we derive the thermodynamic properties of photon--like modes by incorporating the modified dispersion relation arising from the wormhole geometry, focusing on non--interacting particle modes. Remarkably, all calculations are conducted in a fully \textit{analytical} framework.

\end{abstract}

%\keywords{*****}
\maketitle

%%%%%%%%%%%%%%%%%%%%%%%%%%%%%%%%%%%%%%%%%%%%%%%%%%%%%%%%%%%%%%%%%%%%%%%%%%%%%%%%%%%%%%%%%%%%%%%%%%%%%%%%%%%%%%%%%%%%%%%%%%%%%%%%%%%%%%%%%%%%%%%%%%%%%%%%%%%%%%%%%%%%%%%%%%%%%%%%%%%%%%%%%%%%%%%%%%%%%%%%%%%%%%%%%%%%%%%%%%%%%%%%%%%%%%%%%%%%%%%%%%%%%%%%%%%%%%%%%%%%%%%%%%%%%%%%%%%%%%%%%%%%%%%%%%%%%%%%%%%%%%%%%%%%%%%%%%%%%%%%%%%%%%%%%%%%%%%%%%%%%%%%%%%%%%%%%%%%%%%%%%%%%%%%%%%%%%%%%%%%%%%%%%%%%%%%%%%%%%%%%%%%%%%%%

\section{\protect\bigskip Introduction}

In the context of general relativity and its extensions, certain spacetime configurations can support exotic structures, such as wormholes with non--trivial topology. A wormhole can be conceptualized as a tunnel--like passage, linking either two separate universes or two distant, asymptotically flat regions within the same universe. These structures naturally arise as solutions to Einstein’s field equations, with early work on the topic attributed to Flamm \cite{flamm1916beitrage} and later expanded upon by Einstein and Rosen \cite{einstein1935particle}. Their studies laid the foundation for Wheeler’s subsequent exploration of the implications of such solutions \cite{wheeler1955geons}. At that stage, wormholes were largely considered theoretical constructs \cite{ovgun2019exact,Ovgun:2018fnk,Javed:2019qyg,Ovgun:2020yuv,Panyasiripan:2024kyu}, as they did not provide the conditions necessary for traversable spacetimes.

% Ovgun:2018ran,Ovgun:2018uin,Halilsoy:2013iza,Ovgun:2016ijz,Javed:2022fsn,Mustafa:2021vqz,Ovgun:2018oxk

Research has increasingly concentrated on the potential stability and traversability of wormholes, often requiring the presence of exotic matter to ensure their stabilization \cite{visser1995lorentzian}. Beyond the theoretical possibility of faster--than--light travel, wormholes have also attracted attention for their intriguing mathematical characteristics, particularly regarding their relationship with energy conditions \cite{bouhmadi2014wormholes}. In addition, there is ongoing interest in the cosmological ramifications of wormholes, including their possible influence during the early stages of the universe and their role in shaping cosmic structures \cite{cataldo2009evolving}. Investigations into their quasinormal modes have further expanded the understanding of their dynamic properties \cite{b1,b3}.

The principle of observational consistency across all inertial frames relies heavily on the preservation of Lorentz symmetry \cite{verlindearaujo}. This foundational concept, which governs both rotational and boost transformations, underpins both general relativity and the standard model of particle physics. In curved spacetimes, this symmetry holds locally, mirroring the Lorentzian structure of the underlying geometry. Yet, when inertial conditions are no longer satisfied, subtle effects based on direction or velocity can emerge, altering the behavior of particles and wave propagation \cite{STR7,STR2,STR1,amarilo2024gravitational,STR5}.

Symmetry breaking often uncovers fascinating consequences, frequently indicating the emergence of new physical phenomena. In particular, Lorentz symmetry breaking (LSB) introduces a range of unique features \cite{hees2016,liberati2013,tasson2014}, offering significant insights into quantum gravity theories \cite{rovelli2004}. Various theoretical frameworks incorporate the violation of Lorentz invariance, such as closed string theories \cite{New2,New1}, loop quantum gravity \cite{New7,New6}, and noncommutative spacetime models \cite{New9,New8}. Additionally, non--local gravity theories \cite{Modesto:2011kw}, spacetime foam models \cite{New11,New10}, and (chiral) field theories defined on spacetimes with complex topologies \cite{New15,New14} all explore the possibility of Lorentz symmetry violation. Furthermore, Hořava--Lifshitz gravity \cite{New16} and certain cosmological models \cite{sv2,sv1} also consider LSB as a central element in their approach \cite{Kluson:2011rs,Kluson:2010za,Nojiri:2010kx,Nojiri:2010tv,Carloni:2010nx}.

Incorporating LSB into gravitational models presents distinct challenges compared to its inclusion in non--gravitational field theories, where Lorentz violation (LV) terms can be added more straightforwardly. Taking into account flat spacetime, terms such as those from the Carroll--Field--Jackiw model \cite{CFJ} and aether theories \cite{aether} can be seamlessly integrated. For a detailed treatment of all possible minimal LV couplings \cite{colladay1998lorentz}.

A comprehensive framework that thoroughly addresses potential coefficients for Lorentz and CPT violation, including gravitational effects, is encapsulated within the Standard Model Extension (SME) \cite{bluhm2021gravity,Filho:2022yrk,kostelecky2004gravity,araujo2024gravitational,kostelecky2021backgrounds,araujo2024exact,bluhm2005spontaneous,araujo2024exploring,bluhm2023spontaneous,bluhm2008spontaneous}. In its gravitational sector, this framework functions on a Riemann--Cartan manifold, incorporating torsion as a dynamic geometric element alongside the metric \cite{pp7,pp2,pp4}.

Investigating thermal radiation in the framework of LSB offers profound insights into the early Universe's characteristics \cite{araujo2022thermal}. This approach is grounded in the observation that, during the Universe's primordial phase, its size aligns with the scales associated with Lorentz violation \cite{amelino2001testable,kostelecky2011data,lsb1,lsb2,lsb3,lsb4,lsb5,lsb6}. The exploration of LSB's thermal properties was first introduced in \cite{colladay2004statistical}, and since then, a wide range of studies have addressed different scenarios. These include investigations into linearized gravity \cite{aa2021lorentz}, Pospelov and Myers--Pospelov electrodynamics \cite{anacleto2018lorentz,araujo2021thermodynamic}, and both CPT--even and CPT--odd Lorentz--violating (LV) terms \cite{casana2009finite,casana2008lorentz,araujo2021higher,aguirre2021lorentz}. Other research has examined higher--dimensional operators \cite{reis2021thermal,Mariz:2011ed}, bouncing universe models \cite{petrov2021bouncing2}, and Einstein--aether theory \cite{aaa2021thermodynamics}.

However, despite the exploration of modified dispersion relations, no thermal analysis of massless particles has yet been conducted within the context of a bumblebee wormhole scenario. In this manner, this work investigates several phenomena influenced by the gravitational field in a bumblebee gravity framework, with a particular focus on a traversable wormhole for massless particle modes. In this context, the refractive index, group velocity, time delay, modified distances, and interparticle potential are computed, highlighting the possibility of photon--photon interactions due to the wormhole geometry. Additionally, for the interparticle potential, the analysis is extended to massive particle modes, resulting in a combination of modified Yukawa-- and Coulomb--like potentials. These calculations depend on the parameters of the wormhole, particularly the size of the throat. In addition, the Hawking temperature is derived using the trapping horizon method, which shows negative values. Finally, thermodynamic properties for photon--like modes are derived by considering the modified dispersion relation induced by the wormhole geometry, accounting for non--interacting particle modes. In particular, all these calculations are performed \textit{analytically}.

%%%%%%%%%%%%%%%%%%%%%%%%%%%%%%%%%%%%%%%%%%%%%%%%%%%%%%%%%%%%%%%%%%%%%%%%%%%%%%%%%%%%%%%%%%%%%%%%%%%%%%%%%%%%%%%%%%%%%%%%%%%%%%%%%%%%%%%%%%%%%%%%%%%%%%%%%%%%%%%%%%%%%%%%%%%%%%%%%%%%%%%%%%%%%%%%%%%%%%%%%%%%%%%%%%%%%%%%%%%%%%%%%%%%%%%%%%%%%%%%%%%%%%%%%%%%%%%%%%%%%%%%%%%%%%%%%%%%%%%%%%%%%%%%%%%%%%%%%%%%%%%%%%%%%%%%%%%%%%%%%%%%%%%%%%%%%%%%%%%%%%%%%%%%%%%%

\section{The fundamental aspects of the theory}

In this section, we review the exact solution of bumblebee wormhole presented in Ref. \cite{ovgun2019exact}. Let us start with the  following bumblebee action
\begin{equation}
S_{B} = \int dx^{4} \sqrt{-g} \left[\frac{R}{2\kappa} + \frac{1}{2\kappa}\xi B^{\mu}B^{\nu} R_{\mu\nu} - \frac{1}{4} B_{\mu\nu}B^{\mu\nu} - V(B_{\mu}B^{\mu} \pm \Tilde{b}^{2})\right] + \int dx^{4} \mathcal{L}_{m},
\end{equation}
In this framework, \( B_{\mu} \) denotes the bumblebee vector field, while the field strength tensor is given by \( B_{\mu\nu} = \partial_{\mu}B_{\nu} - \partial_{\nu}B_{\mu} \). The parameter \( \xi \) represents the constant that couples the bumblebee field to the curvature in a non--minimal way. For vacuum configurations where \( V(B_{\mu}B^{\mu} \mp \tilde{b}^{2}) = 0 \), the expression \( \tilde{b}^2 = \pm B^{\mu}B_{\mu} = \pm \tilde{b}^{\mu}\tilde{b}_{\mu} \) corresponds to the norm of the non--zero vector associated with the vacuum expectation value \( \langle B^{\mu} \rangle = \tilde{b}^{\mu} \) \cite{kostelecky2004gravity}. Here, \( R \) stands for the scalar curvature, \( g \) is the determinant of the metric tensor, and \( \kappa \) denotes the gravitational constant.

The stress--energy tensor is altered due to the presence of the bumblebee field, taking the form shown below \cite{casana2018exact}:
\begin{equation}
R_{\mu \nu }-\kappa G\left[ T_{\mu \nu }^{M}+T_{\mu \nu }^{B}-\frac{1}{2}%
g_{\mu \nu }\left( T^{M}+T^{B}\right) \right] =0, \label{einstein-1}
\end{equation}%
where \( T^{M} = g^{\mu \nu }T_{\mu \nu }^{M} \), and the stress--energy tensor for the bumblebee field, \( T_{\mu\nu}^{B} \), is expressed as 
\begin{gather}
T_{\mu \nu }^{B}=-B_{\mu \alpha }B_{\nu }^{\alpha }-\frac{1}{4}B_{\alpha
\beta }B^{\alpha \beta }g_{\mu \nu }-Vg_{\mu \nu }+2V^{\prime }B_{\mu
}B_{\nu }+  \nonumber \\
+\frac{\xi }{\kappa }\Big[\frac{1}{2}B^{\alpha }B^{\beta }R_{\alpha \beta
}g_{\mu \nu }-B_{\mu }B^{\alpha }R_{\alpha \nu }-B_{\nu }B^{\alpha
}R_{\alpha \mu }+  \nonumber \\
+\frac{1}{2}\nabla _{\alpha }\nabla _{\mu }(B^{\alpha }B_{\nu })-\frac{1}{2%
}\nabla ^{2}(B_{\mu }B_{\nu })+  \nonumber \\
-\frac{1}{2}g_{\mu \nu }\nabla _{\alpha }\nabla _{\beta }(B^{\alpha
}B^{\beta })\Big]. \label{asdemdtensor}
\end{gather}%

The stress--energy tensor from Eq. \eqref{asdemdtensor}, along with the modified Einstein equation from Eq. \eqref{einstein-1}, can be written explicitly as follows:
\begin{gather}
E_{\mu \nu }^{instein} = R_{\mu \nu }-\kappa \left( T_{\mu \nu }^{M}-\frac{1%
}{2}g_{\mu \nu }T^{M}\right) -\kappa T_{\mu \nu }^{B}-2\kappa g_{\mu \nu }V+ 
\nonumber \\
+\kappa B_{\alpha }B^{\alpha }g_{\mu \nu }V^{\prime }-\frac{\xi }{4}g_{\mu
\nu }\nabla ^{2}(B_{\alpha }B^{\alpha }) + \nonumber \\
-\frac{\xi }{2}g_{\mu \nu }\nabla _{\alpha }\nabla _{\beta }(B^{\alpha
}B^{\beta })=0.  \label{einstein-2}
\end{gather}%
At this stage, the authors in Ref. \cite{ovgun2019exact} adopt a static, spherically symmetric traversable bumblebee wormhole solution, which is expressed in the following form \cite{morris1988wormholes}:
\begin{equation}
\mathrm{d}s^2 = e^{2\Lambda(r)} \mathrm{d}t^{2} - \frac{\mathrm{d}r^{2}}{1 - \frac{b(r)}{r}} - r ^2 \mathrm{d}\theta^{2} - r^2 \sin^{2}\theta \mathrm{d}\varphi^{2}. \label{metric}
\end{equation}
In this case, the redshift function is taken to be zero \((\Lambda = 0)\), and the bumblebee vector \(\tilde{b}_{\mu}\) is aligned with the wormhole shape function \(b(r)\), as described below \cite{ovgun2019exact}:
\begin{equation}
\Tilde{b}_{\mu} = \left(0, \sqrt{\frac{c}{1 - \frac{b(r)}{r}}}, 0,0 \right),
\end{equation}
with \( c \) is a positive constant related to the term responsible for Lorentz violation.

Additionally, as outlined in Ref. \cite{ovgun2019exact}, the isotropic stress-energy tensor can be expressed in terms of a perfect fluid, \( (T^{\mu}_{\nu})^{M} = (\rho, -P, -P, -P) \), with
\begin{eqnarray}
P=w \rho, 
\end{eqnarray}
under the assumption \(\rho \geq 0\). The dimensionless constant \( -\frac{1}{3} < w \leq 1 \) plays a crucial role in maintaining the validity of the energy conditions. By inserting the wormhole metric from Eq. \eqref{metric} into the Einstein equation \eqref{einstein-2}, we obtain the modified Einstein equations that now depend on both the shape function \( b(r) \) and the parameter \( w \) \cite{ovgun2019exact}
\begin{gather}
\label{aedidnfsgthehijn}
G_{t t }= -\kappa \rho {r}^{3}\left(1+3w\right)+\lambda
r\dot{b}(r) -\lambda b(r)=0
,\\ 
G_{r r }=  \kappa \rho {r}^{3}\left(w-1\right)+\left(2+3\lambda\right)r\dot{b}-\left(2+3\lambda\right)b(r)=0
,\\
G_{\theta \theta }= \kappa \rho {r}^{3}\left(w-1\right)+r\dot{b}(r)+\left(2\lambda+1\right)b(r)-2\lambda r =0
,\label{123e123i123n123s2t3dein}
\end{gather}
where \(\lambda = a \xi\) represents the Lorentz symmetry breaking (LSB) parameter, and the dot signifies differentiation with respect to the radial coordinate \(r\). From the system of Eqs. \(\left(\ref{aedidnfsgthehijn}-\ref{123e123i123n123s2t3dein}\right)\), the energy density \(\rho\) can be extracted. Specifically, solving Eq. \eqref{aedidnfsgthehijn} for \(\rho\) yields the following result:
\begin{equation}
\label{hgass}
\rho = \frac{\lambda\left( r \dot{b}(r)  - b(r)\right)  }{\kappa r^{3} (1 + 3w)}.
\end{equation}

Furthermore, the shape function \( b(r) \) can be determined by first multiplying Eq. \eqref{aedidnfsgthehijn} by \((w-1)\) and then adding it to Eq. \eqref{123e123i123n123s2t3dein} multiplied by \((1+3w)\). Additionally, applying the condition at the wormhole throat, \( b(r_0) = r_0 \), results in the following solutions \cite{ovgun2019exact}
\begin{equation}
b(r) = \frac{\lambda r}{\lambda + 1} + \frac{r_{0}}{\lambda + 1} \left(\frac{r_{0}}{r} \right)^{\gamma}, \label{abffr}
\end{equation}
with $\gamma(w,\lambda)=\frac{\lambda (5 w+3)+3 w+1}{\lambda (w-1)+3 w+1}$. Notice that \(\dot{b}(r)\) can be explicitly expressed as follows:
\begin{equation}
\dot{b}(r) = \frac{\lambda}{\lambda + 1}\left[1-\gamma\left(\frac{r_{0}}{r} \right)^{\gamma+1}\right]. \label{12drbf}
\end{equation}

By utilizing Eq. \eqref{abffr} and Eq. \eqref{12drbf}, the energy density in Eq. \eqref{hgass} can be derived in the following form:
\begin{equation}
\rho(r) =  - \frac{2 \lambda r_0^{\gamma+1}r^{-\left(\gamma+3\right)}}{\kappa \left(\lambda (w-1)+3 w+1\right)}.
\end{equation}

%%%%%%%%%%%%%%%%%%%%%%%%%%%%%%%%%%%%%%%%%%%%%%%%%%%%%%%%%%%%%%%%%%%%%%%%%%%%%%%%%%%%%%%%%%%%%%%%%%%%%%%%%%%%%%%%%%%%%%%%%%%%%%%%%%%%%%%%%%%%%%%%%%%%%%%%%%%%%%%%%%%%%%%%%%%%%%%%%%%%%%%%%%%%%%%%%%%%%%%%%%%%%%%%%%%%%%%%%%%%%%%%%%%%%%%%%%%%%%%%%%%%%%%%%%%%%%%%%%%%%%%%%%%%%%%%%%%%%%%%%%%%%%%%%%%%%%%%%%%%%%%%%%%%%%%%%%%%%%%%%%%%%%%%%%%%%%%%%%%%%%%%%%%%%%%%%%%%%%%%%%%%%%%%%%%%%%%%%%%%%%%%%%%%%%%%%%%%%%%%%%%%%%%%%%%%%%%%%%%%%%%%%%%%%%%%%%%%%%%%%%%%%%%%%%%%%%%%%%%%%%%%%%%%%%%%%%%%%%%%%%%%%%%%%%%%%%%%%%%%%%%%%%%%%%%%%%%%%%%%%%%%%%%%%%%%%%%%%%%%%%%%%%%%%%%%%%%%%%%%%%%%%%%%%%%%%%%%%%%%%%%%%%%%%%%%%%%%%%%%%%%%%%%%%%%%%%%%%%%%%%%%%%%%%%%%%%%%%%%%%%%%%%%%%%%%%%%%%%%%%%%%%%%%%%%%%%%%%%%%%%%%%%%%%%%%%%%%%%%%%%%%%%%%%%%%%%%%%%%%

\section{Particle motion in bumblebee wormhole}

\begin{figure}
    \centering
     \includegraphics[scale=0.32]{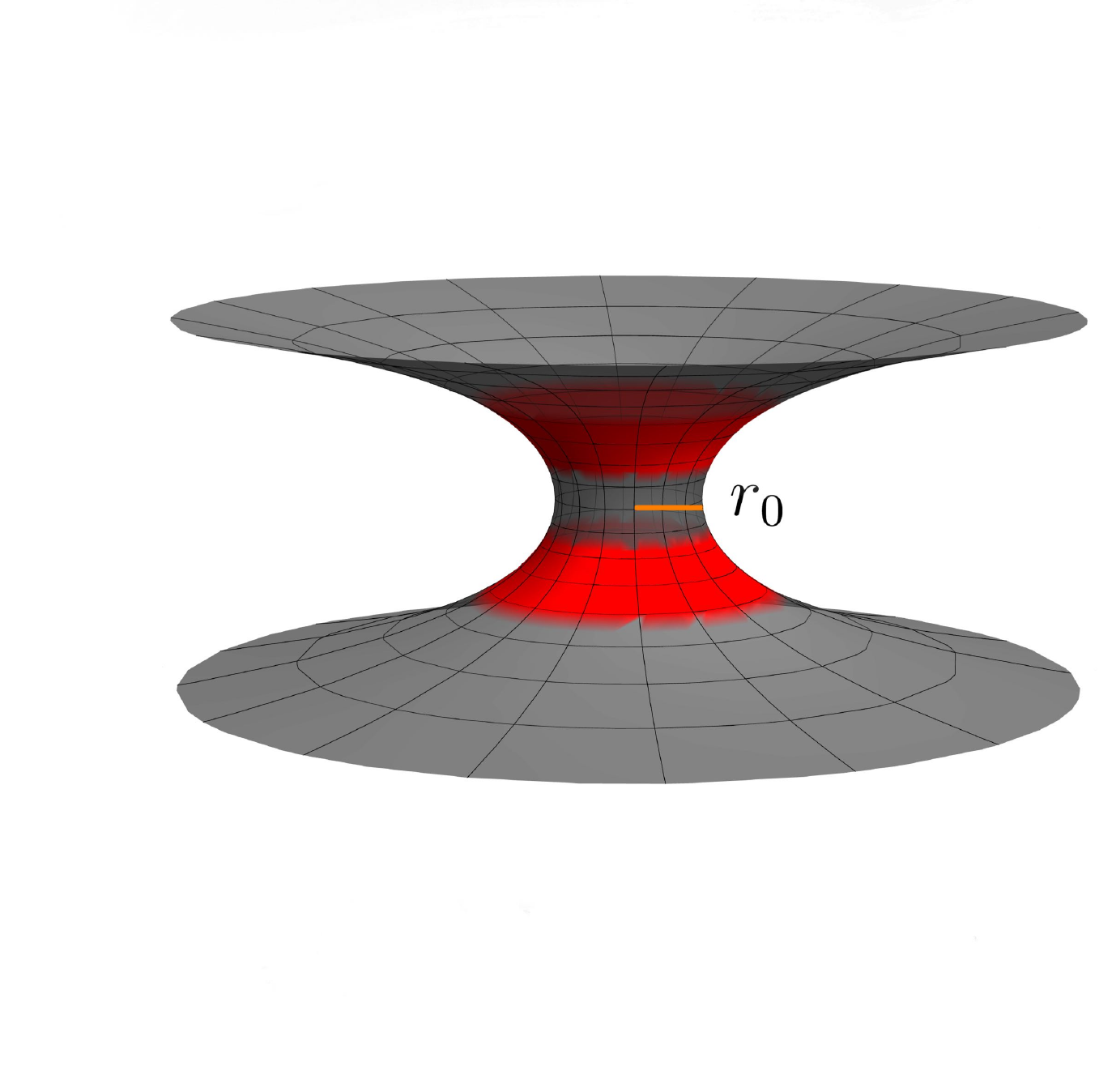}
     \includegraphics[scale=0.43]{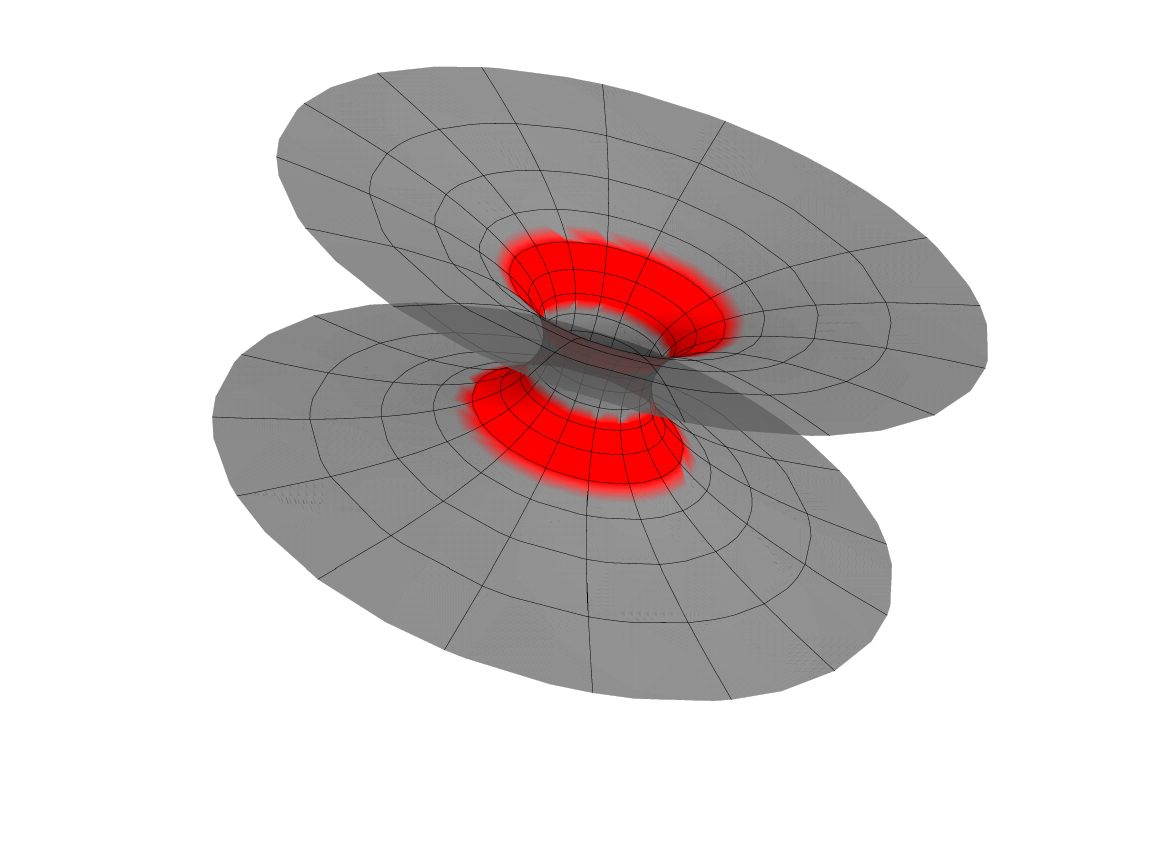}
    \caption{The illustration of the wormhole. In this depiction, $r_{0}$ denotes the throat, while the red zone delineates the region close to it.}
    \label{wormholepics}
\end{figure}

We start by briefly outlining the key aspects of the bumblebee wormhole's geometric structure. Following this, the relationship between the Hamiltonian and momentum for a massive particle within this framework is derived. The thermodynamic behavior is then analyzed in three distinct regions: near the throat, very close to the throat, and far from it. For clarity, Fig. \ref{wormholepics} provides a visual representation. The static, spherically symmetric wormhole metric in isotropic coordinates is expressed as follows:
\ie
\mathrm{d}\mathrm{s}^{2} = \Omega^{2}(r) \,\mathrm{d}t^{2} - \Phi^{-2}(r)\mathrm{d}r^{2} + r^{2}(\mathrm{d}\theta^{2} + \sin^{2}\theta \,\mathrm{d}\varphi^{2}). \label{metric}
\fe
In the case of the bumblebee wormhole, the redshift function \(\Omega\) is expressed as
\ie
\Omega^{2}(r) = - 1,  \label{omegazao}
\fe
and $\Phi$ takes the form
\ie
\Phi^{-2}(r) = - \frac{1}{1 - \frac{b(r)}{r}}.\label{phizao}
\fe
A crucial point to highlight is that the bumblebee wormhole lacks asymptotic flatness. Having outlined the wormhole’s geometry, we now shift focus to the behavior of particles traversing this curved spacetime. The motion of a massive particle is described by the following action:
\ie
S= -m_0 c \int \mathrm{d}{s},
\fe
in which the integral is computed along the particle's worldline. Notice that by using Eq. (\ref{metric}), the corresponding Lagrangian for the particle is expressed as \cite{nandi2016stability}:
\ie
\mathcal{L} \equiv - \ m_{0} c^{2} \Omega \sqrt{ 1 - \frac{v^{2}}{c^{2}}(\Omega\Phi)^{-2} }.
\fe
Likewise, the momentum of the particle in the gravitational field can be written as given below:
\ie
{\bf{p}} = \frac{ (\Omega\Phi^2)^{-1} m_{0} {\bf{v}}}{ \sqrt{1-\frac{v^{2}}{c^{2}}(\Omega\Phi)^{-2} }}.
\fe
In possession of the above expression, and considering the dispersion relation between energy ($E$) and momentum (${\bf{p}}$), $E = \sqrt{(m_0 c^2)^2 + ({\bf{p}}  c)^2}$, we have
\ie
E = \sqrt{(m_0 c^2)^2 + \left(\frac{m_0 \, c \, {\bf{v}} }{\sqrt{1 - \frac{v^2}{c^2}} \, (\Omega \Phi)^{-2}}  \right)^2},
\fe
or simply
\ie
\mathcal{H} = E =  m_{0} c^{2}\Omega \sqrt{1 + \Phi^{2}\frac{p^{2}}{m_{0}^{2}c^{2}}},
\fe
where $\mathcal{H}$ is the Hamiltonian. The de Broglie matter waves can be obtained from the preceding equation below \cite{nandi2016stability}
\ie
\label{asddasd}
\hbar^{2}E^{2} = m^{2}_{0}c^{4} \Omega^{2}\left(1 + \frac{\Phi^2 \hbar^{2}k^{2}}{(m_0 c)^{2}}\right).
\fe
In this context, \(k\) stands for the momentum, and \(m_{0}\) indicates the rest mass. A significant focus of research has been placed on modified dispersion relations (MDRs), especially at the convergence of quantum mechanics and general relativity \cite{amelino2013quantum,ling2006modified}. MDRs are designed to address inconsistencies that emerge from the correlation between quantum effects and gravitational phenomena \cite{smolin2006case,kowalski2005introduction}. These relations are believed to reveal fundamental features about the nature of spacetime, particularly at the Planck scale, where quantum mechanics and gravitational effects become prominent \cite{mattingly2005modern,girelli2009emergence}. Furthermore, from an observational standpoint, these relations have the potential to influence high--energy processes, such as gamma--ray bursts or the propagation of ultra--high--energy cosmic rays \cite{jacob2008lorentz,galaverni2008lorentz}.

%%%%%%%%%%%%%%%%%%%%%%%%%%%%%%%%%%%%%%%%%%%%%%%%%%%%%%%%%%%%%%%%%%%%%%%%%%%%%%%%%%%%%%%%%%%%%%%%%%%%%%%%%%%%%%%%%%%%%%%%%%%%%%%%%%%%%%%%%%%%%%%%%%%%%%%%%%%%%%%%%%%%%%%%%%%%%%%%%%%%%%%%%%%%%%%%%%%%%%%%%%%%%%%%%%%%%%%%%%%%%%%%%%%%%%%%%%%%%%%%%%%%%%%%%%%%%%%%%%%%%%%%%%%%%%%%%%%%%%%%%%%%%%%%%%%%%%%%%%%%%%%%%%%%%%%%%%%%%%%%%%%%%%%%%%%%%%%%%%%%%%%%%%%%%%%%

\section{General features }

In this section, we explore the impact of the bumblebee wormhole’s modified dispersion relations on key physical properties. Specifically, we compute the refractive index, group velocity, modified distances, time delay, and interparticle potential, all derived from the wormhole's altered spacetime geometry. This analysis aims to explore how these modifications influence the behavior of light--like particles and their interactions within this framework.

%%%%%%%%%%%%%%%%%%%%%%%%%%%%%%%%%%%%%%%%%%%%%%%%%%%%%%%%%%%%%%%%%%%%%%%%%%%%%%%%%%%%%%%%

\subsection{Index of refraction}

\begin{figure}
    \centering
     \includegraphics[scale=0.5]{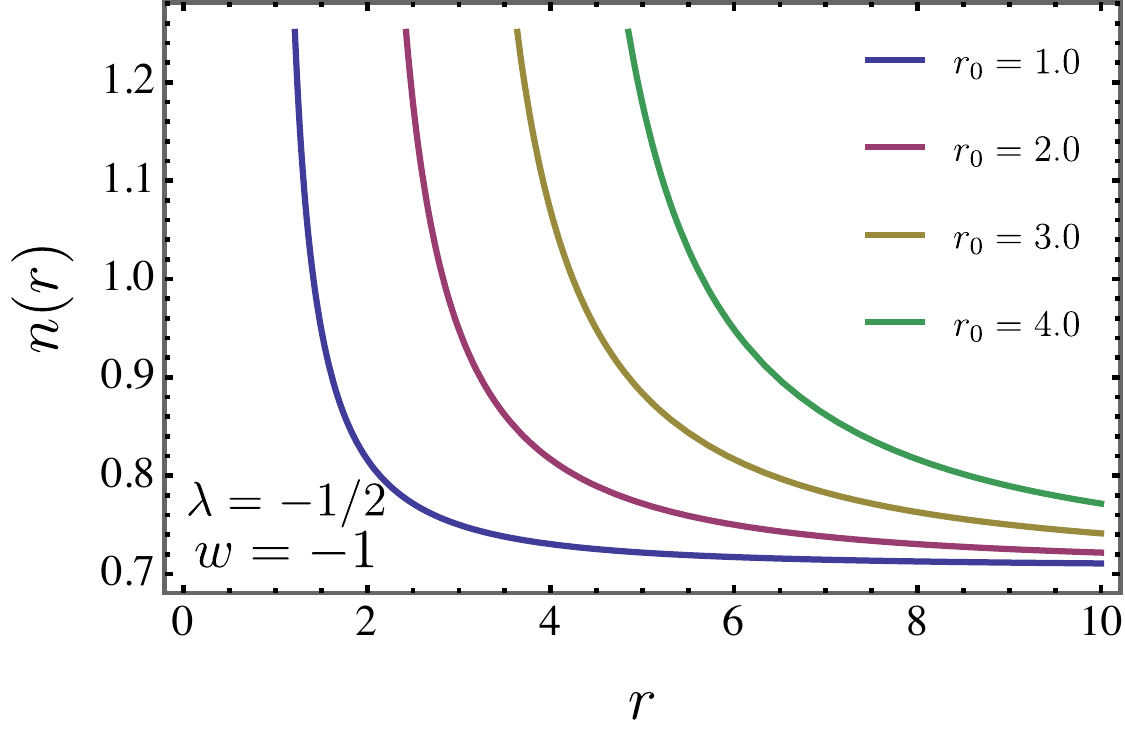}
    \caption{The index of refraction as a function of \( r \) for different configurations of the wormhole's throat, \( r_0 \).}
    \label{index}
\end{figure}

Straightforwardly, we write the index of refraction as follows
\ie
n(r) = \frac{1}{\Phi(r)  \Omega(r) },
\fe
and since we are dealing with the bumblebee wormhole, we have
\ie
n(r) = \frac{1}{\sqrt{\frac{r-r_{0} \left(\frac{r_{0}}{r}\right)^{\frac{3 \lambda +(5 \lambda +3) w+1}{-\lambda +(\lambda +3) w+1}}}{(\lambda +1) r}}}.
\fe
This quantity is shown in Fig. \ref{index}. It is evident that as \( r \) decreases, the index of refraction \( n(r) \) increases rapidly and exhibits asymptotic behavior as it approaches \( r_0 \).

%%%%%%%%%%%%%%%%%%%%%%%%%%%%%%%%%%%%%%%%%%%%%%%%%%%%%%%%%%%%%%%%%%%%%%%%%%%%%%%%%%%%%%%%

\subsection{Group velocity}

\begin{figure}
    \centering
     \includegraphics[scale=0.5]{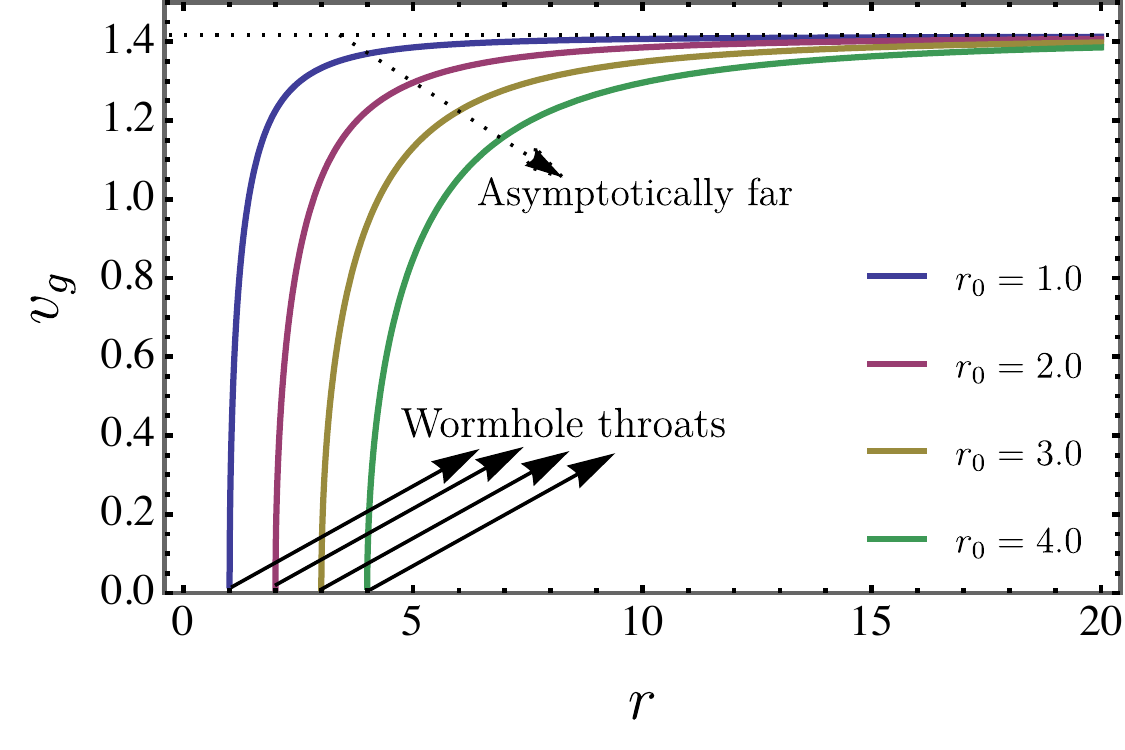}
    \caption{The group velocity as a function of $r$ for a variety of configurations of the wormhole's throat $r_{0}$.}
    \label{vg}
\end{figure}

The group velocity $v_{g}$ is the speed at which a wave packet (and therefore the particle) travels, and is given by the derivative of the energy with respect to the momentum
\ie
v_{g} = \frac{\mathrm{d}E}{\mathrm{d}k} = \frac{k \left(r-r_{0} \left(\frac{r_{0}}{r}\right)^{\frac{3 \lambda +(5 \lambda +3) w+1}{-\lambda +(\lambda +3) w+1}}\right)}{(\lambda +1) r \sqrt{\frac{k^2 \left(r-r_{0} \left(\frac{r_{0}}{r}\right)^{\frac{3 \lambda +5 \lambda  w+3 w+1}{\lambda  (w-1)+3 w+1}}\right)}{(\lambda +1) r}}}.
\fe
Here, we clearly have the wormhole`s parameter dependency, meaning that photons with different distances travel at slightly different speeds. To better comprehend such phenomena, let us consider a position--dependent configuration to the system. It is worth mentioning that, unless stated otherwise, we shall consider throughout the paper $w=-1$, and $\lambda=-1/2$ for accomplish the calculations \cite{oliveira2019quasinormal}. In other words, we present Fig. \ref{vg}  to facilitate our interpretation, setting $k=1$ for the sake of simplicity. Notice that $v_{g} \to 0$ if $r \to r_{0}$; also, if we consider $r \to \infty $, the curves tend a particular asymptotic value, i.e., $\lim\limits_{r \to \infty}v_{g}$.
In this regard, such a limit reads 
\ie
\lim\limits_{r \to \infty}v_{g} = \sqrt{2}.
\fe
As it can be verified, even in the asymptotically far limit, the result do not recover the usual case, i.e., $v_{g}=1$. This result is rather expected, since the bumblebee wormhole geometry is not  asymptotically flat  -- as argued in Ref. \cite{ovgun2019exact}.

%%%%%%%%%%%%%%%%%%%%%%%%%%%%%%%%%%%%%%%%%%%%%%%%%%%%%%%%%%%%%%%%%%%%%%%%%%%%%%%%%%%%%%%%

\subsection{Time delay}

\begin{figure}
    \centering
     \includegraphics[scale=0.5]{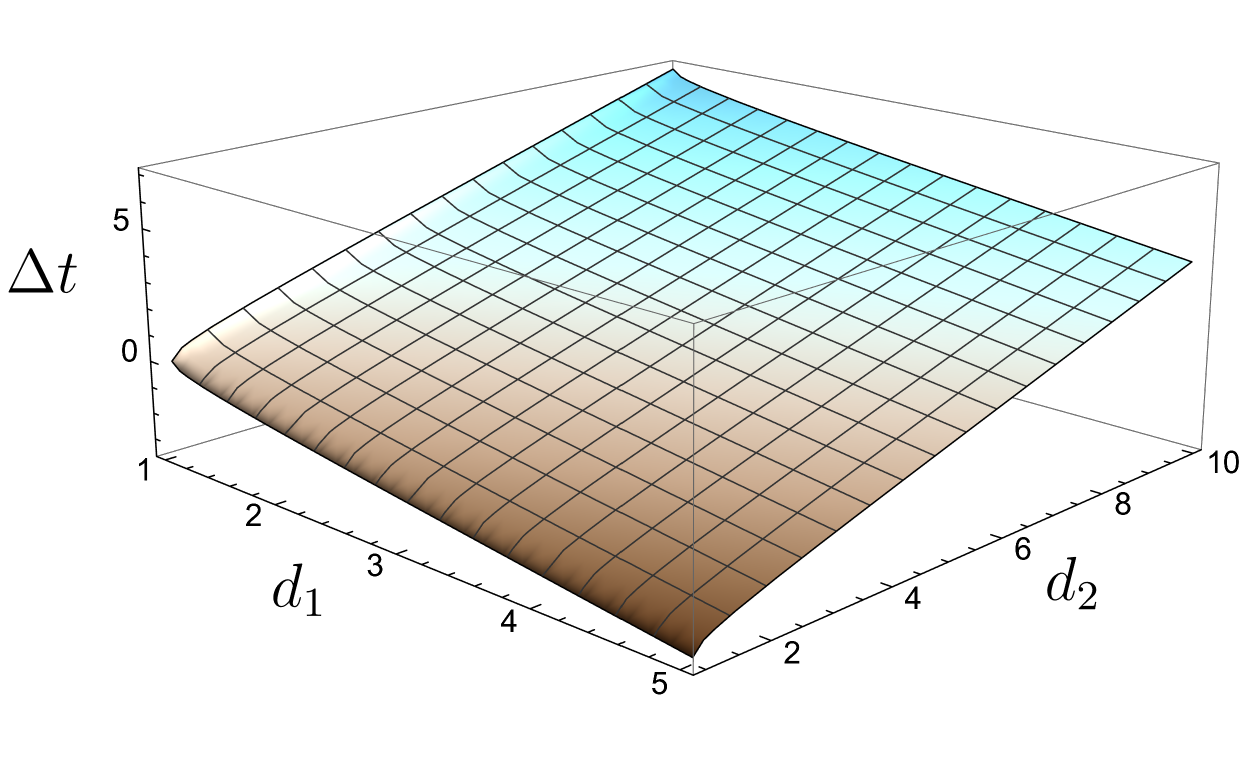}
    \caption{The time delay is represented for two distances, i.e., $d_{1}$ and $d_{2}$.}
    \label{timedelay}
\end{figure}

The time delay $\Delta t$ due to the position--dependent speed of light can be calculated by comparing the travel times of photons with different energies.
In this regard, the travel time $t$ for a photon from a source at a distance $d$ of an observer $\mathcal{O}$ is given by
\ie
\Delta t (d) = t(d_{2}) - t(d_{1}) = \int_{d_{1}}^{d_{2}} \frac{\mathrm{d}r}{v_{g}} = \frac{ \left(\sqrt{d_{2}^2-r_{0}^2} - \sqrt{d_{1}^2-r_{0}^2}\right)}{\sqrt{2} }.
\fe
Remarkably, above result does not depend on the momentum $k$. In Fig. \ref{timedelay}, we show the behavior of the time delay $\Delta t$ for two photon--like particle in $d_{2}$ and $d_{1}$.
Now, let us propose an astrophysical application. Assuming that
$r_0  = 10^6 \, \text{m}$, $
d_1  \approx 9.461 \times 10^{15} \, \text{m} \, (\text{1 light--year})$, 
$d_2  \approx 9.470 \times 10^{15} \, \text{m} \, (\text{1.001 light--years})$, so that the time delay in meters is approximately:
$
\Delta t_{\text{meters}} \approx \frac{1}{\sqrt{2}} \times (9 \times 10^{12}) \, \text{m} \approx 6.36 \times 10^{12} \, \text{m} $. Converting to seconds: $
\Delta t_{\text{seconds}} \approx \frac{6.36 \times 10^{12} \, \text{m}}{3 \times 10^8 \, \text{m/s}} \approx 2.12 \times 10^4 \, \text{seconds} \approx 5.9 \, \text{hours}
$. Therefore, under the influence of the wormhole, an observer \(\mathcal{O}\) in this scenario would measure a time delay of approximately \( 5.9 \, \text{hours} \).

%%%%%%%%%%%%%%%%%%%%%%%%%%%%%%%%%%%%%%%%%%%%%%%%%%%%%%%%%%%%%%%%%%%%%%%%%%%%%%%%%%%%%%%%

\subsection{Modified distances}

Understanding exotic spacetime geometries, like bumblebee wormholes, requires analyzing how these structures alter observable distances. By examining the modified dispersion relation and metric, we can determine how comoving, luminosity, and angular diameter distances are affected. These calculations help predict changes in the apparent size and brightness of distant objects for instance. In this sense, the comoving distance \( D_c^{\text{modified}} \):
\ie
D_c^{\text{modified}} = \int_{r_1}^{r_2} \frac{\mathrm{d}r}{\sqrt{\frac{r-r_{0} \left(\frac{r_{0}}{r}\right)^{\frac{3 \lambda +(5 \lambda +3) w+1}{-\lambda +(\lambda +3) w+1}}}{(\lambda +1) r}}} = \frac{r_{2} \sqrt{1-\frac{r_{0}^2}{r_{2}^2}}}{\sqrt{2}}-\frac{r_{1} \sqrt{1-\frac{r_{0}^2}{r_{1}^2}}}{\sqrt{2}}.
\fe

Furthermore, the luminosity distance \( D_L^{\text{modified}} \):
\[
D_L^{\text{modified}} = (1+z) D_c^{\text{modified}},
\]
where, \( z \) denotes the redshift of a celestial object. In other words, it helps determine the object's distance and the amount the universe has expanded since the light was emitted. Finally, the angular diameter distance \( D_A^{\text{modified}} \) reads:
\[
D_A^{\text{modified}} = \frac{D_c^{\text{modified}}}{1+z}.
\]

%%%%%%%%%%%%%%%%%%%%%%%%%%%%%%%%%%%%%%%%%%%%%%%%%%%%%%%%%%%%%%%%%%%%%%%%%%%%%%%%%%%%%%%%

\subsection{Interparticle potential}

In this section, in order to calculate the interparticle potential $V(r)$, here, we shall use the method of the Green's function method. Particularlly, we consider dispersion relation of the pole of the propagator displayed in Eq. (\ref{asddasd}), which allows us to compute $V(r)$ for both massive and massless particles. In other words, we write
\ie
G(k) = \frac{1}{ \frac{m_{0}^{2} c^{4} \Omega(r)^{2}}{\hbar^{2}}   + \Phi(r)^{2} \Omega(r)^{2} k^{2} } = \frac{1}{\alpha^{2} + \beta^{2} k^{2}},
\fe
where $\alpha^{2} \equiv m_{0}^{2} c^{4} \Omega(r)^{2}/\hbar^{2}$ and $ \beta^{2} \equiv  \Phi(r)^{2} \Omega(r)^{2}$. To obtain the interparticle potential 
$V(r)$, we need to Fourier transform the Green's function $G(k)$ from momentum space to position space. Thereby, it reads \cite{blackledge2005digital,gradshteyn2014table}
\ie
\begin{split}
V(r) &= \int \frac{\mathrm{d}^{3}k}{(2\pi)^{3}} e^{i{\bf{k}} \cdot {\bf{r}}} G(k)\\
& = \frac{1}{(2\pi)^3} \int_{0}^{\infty} \mathrm{d}k \, k^2 \int_{0}^{\pi} \int^{2\pi}_{0} \mathrm{d} \varphi \, \mathrm{d}\theta \, \sin(\theta) \, e^{i k r \cos(\theta)} \, G(k) \\
& = \frac{1}{2 r \pi^2} \int_{0}^{\infty} \mathrm{d} k \, k \sin(kr) \, G(k) \\
&= \frac{1}{2 r \pi^2} \int_{0}^{\infty} \mathrm{d} k \, k \sin(kr) \, \left[   \frac{1}{\alpha^{2} + \beta^{2} k^{2}} \right] \\
& = \frac{e^{-\frac{\alpha r}{\beta} }}{4\pi r \beta^{2}} = \frac{r^2 e^{-\frac{m_{0} r}{\sqrt{2-\frac{2 r_{0}^2}{r^2}}}}}{8 \pi  \left(r^2-r_{0}^2\right)}.
\end{split}
\fe

\begin{figure}
    \centering
     \includegraphics[scale=0.5]{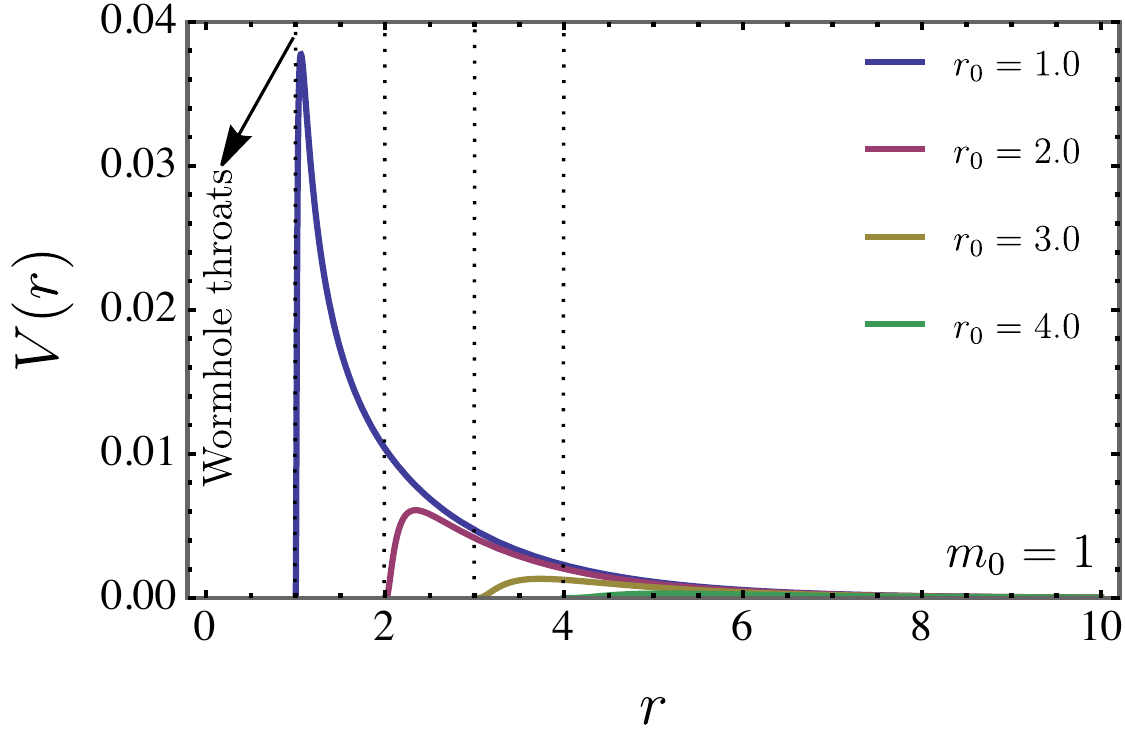}
    \caption{The representation of the interparticle potential $V(r)$ as a function of $r$, taking into account distinc values of $r_{0}$.}
    \label{interparticlepotential}
\end{figure}

\begin{figure}
    \centering
     \includegraphics[scale=0.5]{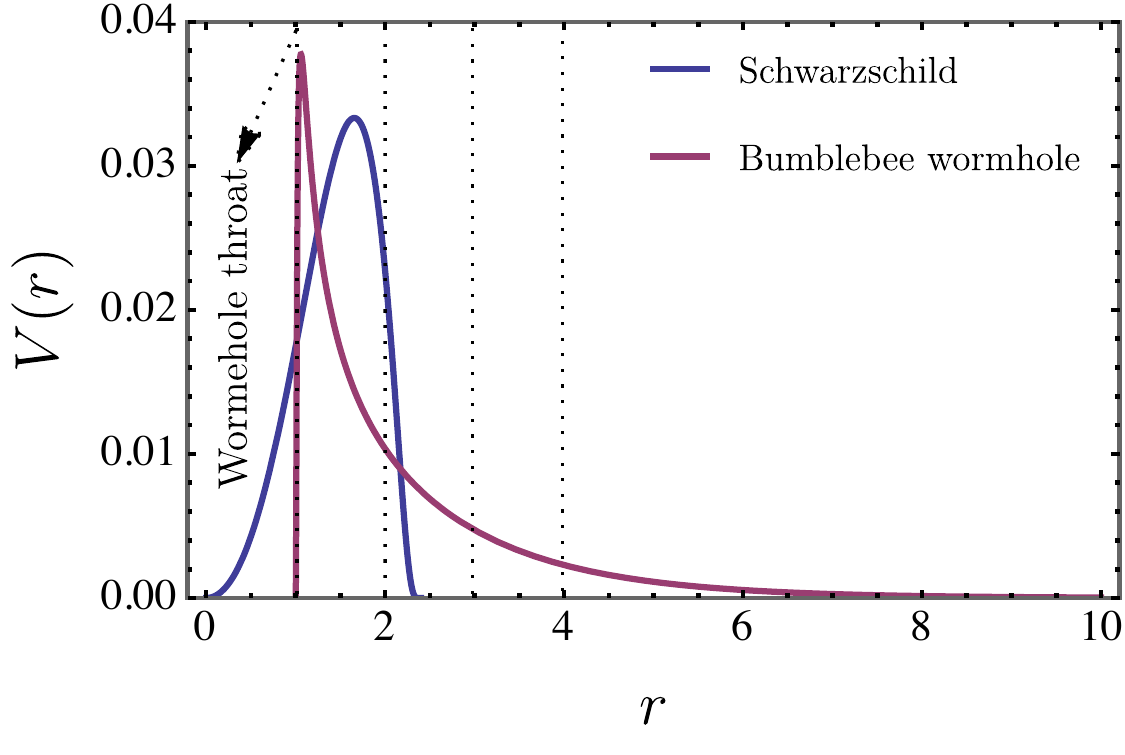}
    \caption{The comparison of the interparticle potential \(V(r)\) for the Schwarzschild case and the bumblebee wormhole, plotted as a function of \(r\).}
    \label{interparticlepotentialcomparison}
\end{figure}

It is important to note that the presence of wormhole geometry naturally leads to a ``combination'' of Yukawa-- and Coulomb--like interactions for the massive particles. The magnitude of the interactions is determined by the wormhole's parameters, such as the throat. Additionally, the potential \( V(r) \) approaches zero as \( r \) tends to both infinity, \(\lim\limits_{r \to \infty} V(r) = 0\), and zero, \(\lim\limits_{r \to 0} V(r) = 0\).
Fig. \ref{interparticlepotential} displays the interparticle potential \( V(r) \) as a function of \( r \) for various values of the wormhole's throat \( r_{0} \). In other words, this potential
$V(r)$ likely models the interaction energy of a test particle or field in the spacetime of a bumblebee wormhole. It captures both the local effects near the wormhole throat, where the geometry causes significant changes in the potential, and the asymptotic behavior far from the throat, where the influence of the wormhole wanes. The presence of the bumblebee field modifies the potential in a way that reflects the underlying Lorentz symmetry breaking in the theory. To facilitate a comparison with the standard Schwarzschild case, we present Fig. \ref{interparticlepotentialcomparison}.

And what about the massless case? To ensure this point, we consider $m_{0} \to 0$, which yields
\ie
V_{0}(r) = \frac{r^2 }{8 \pi  \left(r^2-r_{0}^2\right)},
\fe
where $V_{0}(r)$ represents $V(r)$ considering $m_{0} \to 0$.Note that this results in a modified Coulomb--like interaction. In other words, the wormhole geometry naturally leads to the possibility of photon--photon interactions.
To further understand the behavior of \( V_{0}(r) \), we analyze the potential in different limiting cases. As \( r \) approaches the throat \( r_0 \), \( \lim\limits_{r \to r_{0}} V_{0}(r) \) results in an indeterminate form, indicating a divergent behavior near the wormhole throat, where the spacetime geometry has a significant impact. In contrast, as \( r \) approaches zero, \( \lim\limits_{r \to 0} V_{0}(r) \), the potential smoothly tends to zero, with no singularity, suggesting that the influence of the wormhole diminishes near the origin. Finally, in the asymptotic limit \( r \to \infty \), \( \lim\limits_{r \to \infty} V_{0}(r) \) yields a constant value of \( 1/8\pi \). This nonzero asymptotic value indicates that the wormhole generates a long--range potential field extending throughout the surrounding spacetime. Unlike typical gravitational fields that decay to zero at infinity, the potential $V_{0}(r)$ remains nonzero, converging to a constant value. As a result, even at vast distances, massless particles continue to be influenced by the wormhole. However, this influence is weak and uniform, representing a subtle residual effect of the wormhole's presence.

\begin{figure}
    \centering
     \includegraphics[scale=0.5]{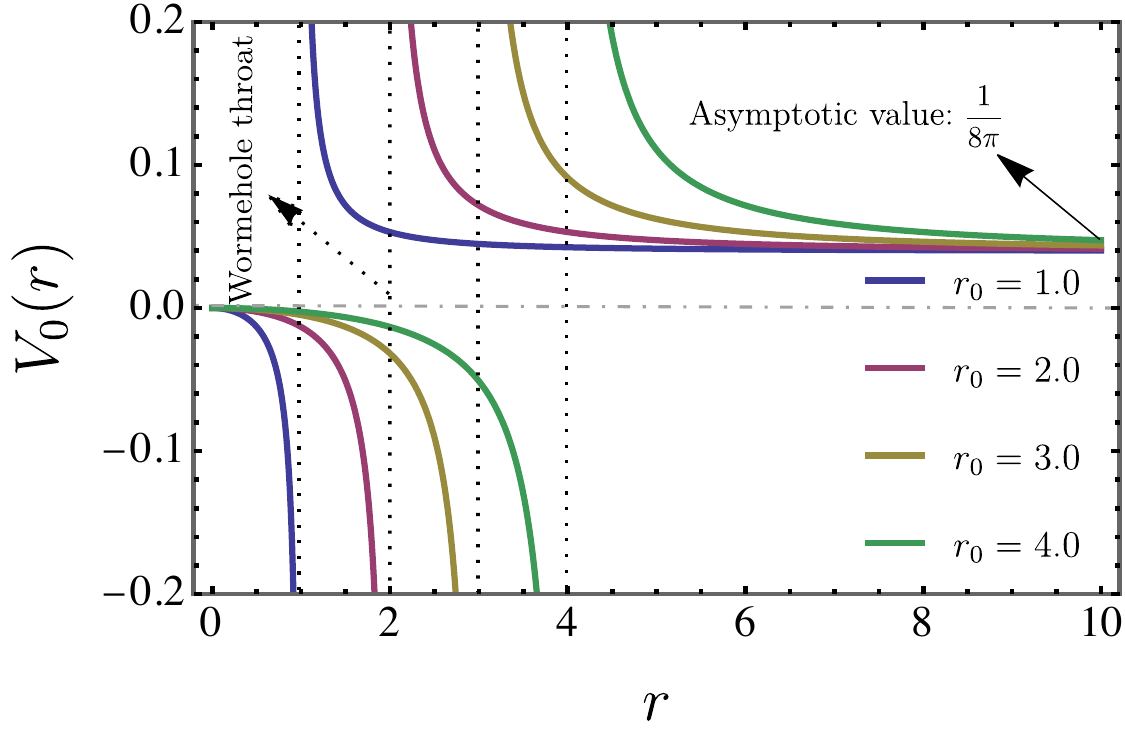}
    \caption{The representation of the interparticle potential $V_{0}(r)$ as a function of $r$, taking into account distinc values of $r_{0}$.}
    \label{interparticlepotentialmassless}
\end{figure}

To corroborate our outputs, in Fig. \ref{interparticlepotentialmassless}, we present the behavior of it. In general lines, we can directly confirm that the plots diverge at $r_{0}$ and tend to a constant value, namely $1/8\pi$, when $  \lim\limits_{r \to \infty} V_{0}(r)$.

%%%%%%%%%%%%%%%%%%%%%%%%%%%%%%%%%%%%%%%%%%%%%%%%%%%%%%%%%%%%%%%%%%%%%%%%%%%%%%%%%%%%%%%%%%%%%%%%%%%%%%%%%%%%%%%%%%%%%%%%%%%%%%%%%%%%%%%%%%%%%%%%%%%%%%%%%%%%%%%%%%%%%%%%%%%%%%%%%%%%%%%%%%%%%%%%%%%%%%%%%%%%%%%%%%%%%%%%%%%%%%%%%%%%%%%%%%%%%%%%%%%%%%%%%%%%%%%%%%%%%%%%%%%%%%%%%%%%%%%%%%%%%%%%%%%%%%%%%%%%%%%%%%%%%%%%%%%%%%%%%%%%%%%%%%%%%%%%%%%%%%%%%%%%%%%%%%%%%%%%%%%%%%%%%%%%%%%%%%%%%%%%%%%%%%%%%%%%%%%%%%%%%%%%%%%%%%%%%%%%%%%%%%%%%%%%%%%%%%%%%%%%%%%%%%%%%%%%%%%%%%%%%%%%%%%%%%%%%%%%%%%%%%%%%%%%%%%%%%%%%%%%%%%%%%%%%%%%%%%%%%%%%%%%%%%%%%%%%%%%%%%%%%%%%%%%%%%%%%%%%%%%%%%%%%%%%%%%%%%%%%%%%%%%%%%%%%%%%%%%%%%%%%%%%%%%%%%%%%%%%%%%%%%%%%%%%%%%%%%%%%%%%%%%%%%%%%%%%%%%%%%%%%%%%%%%%%%%%%%%%%%%%%%%%%%%%%%%%%%%%%%%%%%%%%%%%%%%%%%%%%%%%%%%%%%%%%%%%%%%%%%%%%%%%%%%%%%%%%%%%%%%%%%%%%%%%%%%%%%%%%%%%%%%%%%%%%%%%%%%%%%%%%%%%%%%%%%%%%%%%%%%%%%%%%%%

\section{Thermodynamics via geometry: trapping horizon and temperature}

Wormholes exhibit thermodynamic properties similar to black holes when local physical quantities are used. Since traversable wormholes lack an event horizon, we use the trapping horizon instead \cite{Martin-Moruno:2009rpi,Gonzalez-Diaz:2004clm,Ovgun:2015taa,Rehman:2020myc}. Here, we consider a dynamical wormhole within a cosmological background, effectively generalizing the Morris–-Thorne wormhole to a time--dependent setting 
\ie \label{whdy}
\mathrm{d} s^{2} = - \mathrm{d} t^{2} + a^{2}(t)\left[\frac{\mathrm{d} r^{2}}{1-\frac{b(r)}{r}}+r^{2} \mathrm{d} \Omega^{2}\right].
\fe
The throat of the wormhole is located at the minimum radius \( r = r_0 \), connecting two regions defined by \( r_0 < r < r_a \), with \( r_a \) representing the radius of the wormhole mouth; as \( r \) approaches infinity, the metric becomes flat. The dimensionless parameter \( a(t) \), known as the scaling factor of the universe, describes how the universe is expanding; since the expansion rate is increasing over time, this implies that \( \ddot{a}(t) > 0 \) or that \( \dot{a}(t) \) is an increasing function of time (here, an overdot denotes a time derivative).

We obtain the null coordinates for the above metric by $x^{+} = t + r^{*}$, $
x^{-} = t - r^{*},$ with
$\frac{\mathrm{d} r}{\mathrm{d} r_{*}}=\sqrt{-\frac{g_{00}}{g_{r r}}}=\frac{1}{a(t)} \sqrt{1-\frac{b(r)}{r}}.$ Outgoing and ingoing radiation correspond to \( x^{+} \) and \( x^{-} \), respectively. Utilizing above equations, Eq. (\ref{whdy}) can be rewritten as
\ie
\mathrm{d} s^{2} = 2 g_{+-} \mathrm{d} x^{+} \mathrm{d} x^{-}+R^{2} \mathrm{d} \Omega^{2}.
\fe

Considering  $R$  and  $g_{+-} = -1/2$  as functions of the null coordinates  $\left( x^{+}, x^{-} \right)$ , where  $R = a(t) r$  is known as the areal radius and  $\mathrm{d}\Omega^{2}$  denotes the metric for the unit 2--sphere, we define the expansions as \( \Theta_{\pm} = \dfrac{2}{R} \partial_{\pm} R \). They indicate whether the light rays are expanding ( $\Theta > 0 $) or contracting ($ \Theta < 0 )$, or equivalently, whether the area of the sphere increases or decreases along the null directions.

Since the Killing vector is not applicable in non--stationary spherically symmetric spacetimes, we use the Kodama vector $K$  instead. In null coordinates, it is given by
\ie
K=-g^{+-}\left(\partial_{+} R \partial_{-}-\partial_{-} R \partial_{+}\right)
\fe
applying this to the spacetime described by equation \ref{whdy}, the Kodama vector in covariant form becomes
\ie
K_{ \pm}=-\frac{1}{2}\left( \pm \dot{R}+\sqrt{1-\frac{a b}{R}}\right).
\fe
The norm of the Kodama vector  $K$  is given by
\ie
\|K\|^{2}=\frac{2 E}{R} - 1.
\fe
Importantly, this norm becomes zero on the trapping horizon \(R_{h}\), that is,   \(\|K\|^{2}=0\). In spherically symmetric spacetimes, the active gravitational energy is represented by the Misner--Sharp energy, which reduces to the Newtonian mass in the Newtonian limit for a perfect fluid and yields the Schwarzschild energy in vacuum. At null and spatial infinity, it corresponds to the Bondi--Sachs energy \(E_{B S}\), and the Arnowitt--Deser--Misner energy \(E_{A D M}\), respectively. The Misner--Sharp energy can be expressed as
$
E=\frac{R}{2}\left(1-2 g^{+-} \partial_{+} R \partial_{-} R\right)
$, hence one calculates

\ie
E=\frac{R}{2}\left[\dot{R}^{2}+\frac{a b}{R}\right].
\fe

On a trapping horizon, this expression simplifies to \( E = \dfrac{R_h}{2} \). Using the specified metric, the surface gravity on the trapping horizon becomes
\ie
\kappa=-\frac{\ddot{R_{h}}}{2}+\frac{1}{4 R_{h}^{2}}\left(a b-b^{\prime} R_{h}\right).
\fe

We can also express the surface gravity on a trapping horizon as
$\kappa=\frac{1}{2} g^{a b} \partial_{a} \partial_{b} R
$. From this expression, we observe that\(\kappa<0, \kappa=0\) and \(\kappa>0\) correspond to inner, degenerate, and outer trapping horizons, respectively. The Hawking temperature is given by  \(T=-\kappa_{h} / 2 \pi\), which in our case evaluates to:
\ie
T=-\frac{\left.\kappa\right|_{h}}{2 \pi}=-\frac{1}{2 \pi}\left[-\frac{\ddot{R_{h}}}{2}+\frac{a b-R_{h} b^{\prime}}{4 R_{h}^{2}}\right].
\fe
For the outer trapping horizon, the Hawking temperature is negative because $\kappa_h > 0$. This implies that particles emerging from a wormhole exhibit properties similar to those of phantom energy, as this energy is associated with negative temperature \cite{Martin-Moruno:2009rpi,Gonzalez-Diaz:2004clm,Ovgun:2015taa,Rehman:2020myc}. 
Different cases are now analyzed using specific values of shape functions and a particular cosmological model. The scale factor is taken as \(a(t)=a_{0} t^{n}\) where \(a_{0}\) and 
$n = \frac{2}{3(1+\omega)}$ are constants. $n$ depends on the equation of state parameter $w$, which determines the behavior of the universe’s expansion: For a radiation-dominated universe ($w = \frac{1}{3}$): $n = \frac{1}{2}$. For a matter-dominated universe ($w = 0$): $n = \frac{2}{3}$. For a  quintessence  universe ($-1< w< 0 $): for example $w=-2/3$ and $n=2$. In the static case $(a(t) = 1)$, the wormhole throat is characterized by a bifurcating trapping horizon, as  $R_h = r_0$ and the Kodama vector is observed to take the form
\ie
K_{ \pm}=
-\frac{1}{2} \sqrt{\frac{r - r_0 \left(\frac{r_0}{r}\right)^{\frac{5 \lambda w + 3 \lambda + 3 w + 1}{\lambda w - \lambda + 3 w + 1}}}{r (\lambda + 1)}},
\fe
and, when evaluated on the trapping horizon, the Hawking temperature becomes

\begin{eqnarray} \label{Tst}
T &=&
-\frac{3w + 1}{4\pi R_h \left( \lambda w - \lambda + 3w + 1 \right)}.\end{eqnarray}

\begin{figure}
    \centering
     \includegraphics[scale=0.5]{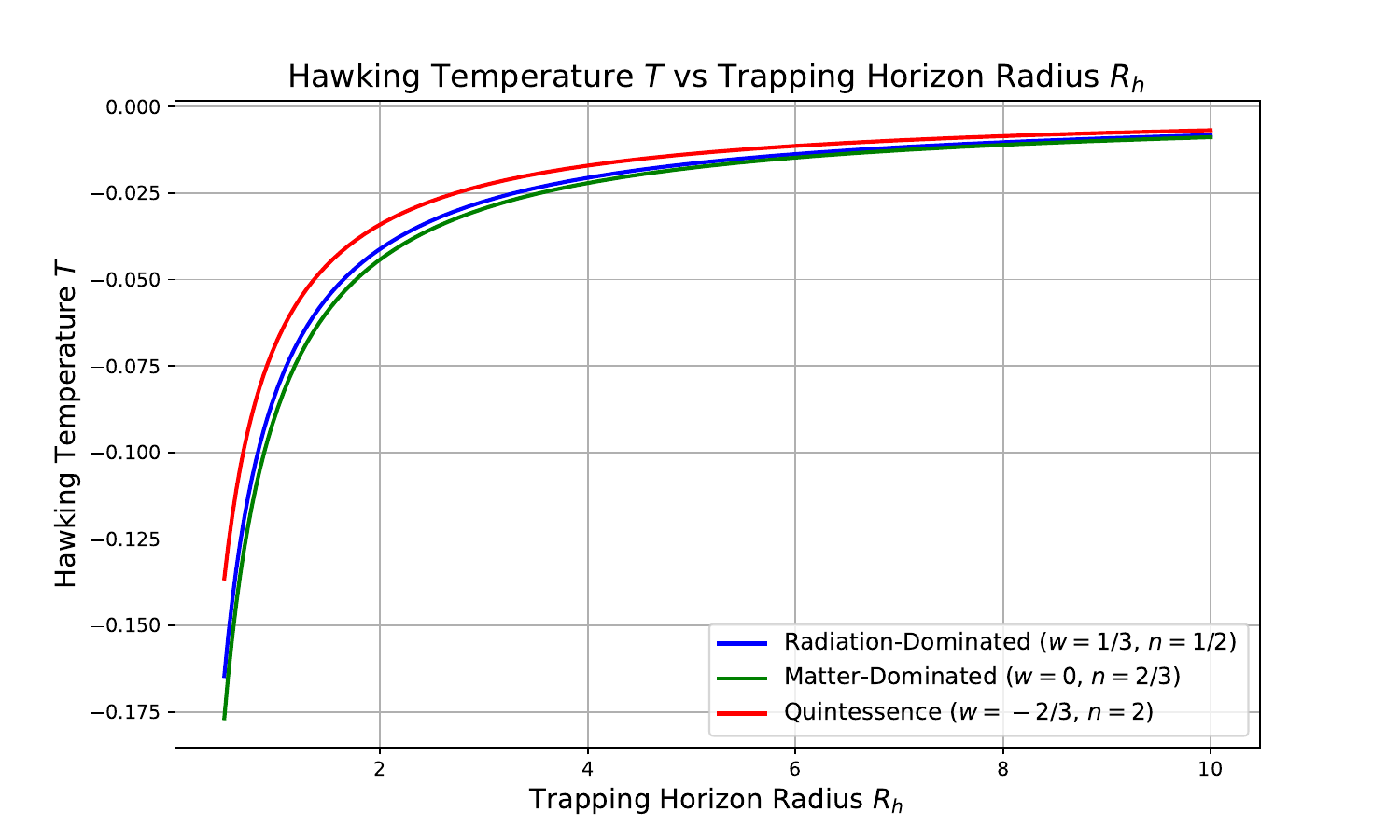}
    \caption{Hawking temperature $T$ as a function of the trapping horizon radius $R_h$, for three cosmological scenarios: radiation--dominated ($w = 1/3$, $n = 1/2$), matter--dominated ($w = 0$, $n = 2/3$), and quintessence ($w = -2/3$, $n = 2$).}
    \label{figurestatic}
\end{figure}

The Fig.  \ref{figurestatic} shows the relationship between the Hawking temperature $T$ and the wormhole radius $R_h=r_0$  in Eq. \ref{Tst} for different cosmological models characterized by the equation of state parameter  $w$ , specifically for  $\lambda = 0.1$. It demonstrates that in radiation-dominated ($ w = \frac{1}{3} $) and matter--dominated ( $w = 0 $) universes, the Hawking temperature inversely correlates with the wormhole radius, indicating that smaller wormholes are hotter and emit more radiation. In contrast, the quintessence scenario ($ w = -\frac{2}{3}$ ) shows a distinct behavior due to the effects of dark energy, with the temperature potentially increasing with the radius.

%For $n=2$, the scale factor is associated with a Phantom-Dominated universe. For $n=2 / 3$, the scale factor is associated with a matter-dominated universe, whereas for $(n=1 / 2$, it represents a radiation-dominated universe.

%Using this scale factor we discuss three cases for different expressions of the shape function.

%The geometry of the wormhole is characterized by the shape function \( b(r) \), which describes its structure as observed from the embedding space in coordinates \( (Z, r, \phi) \), where the two-dimensional surface
%\ieZ(r)= \pm \int\left(\frac{r}{b(r)}-1\right)^{-1 / 2} \mathrm{d} r\fe
%The geometry is identical to that of the two-dimensional surface defined by \( \theta = \pi / 2 \) and \( t = \text{constant} \) in the metric. The function \( Z(r) \), known as the embedding function, describes this geometry. When the graph of Eq.~(2.2) is revolved around the axis of rotation—the \( Z \)-axis—it yields the shape of the wormhole~\cite{23}. This shape function satisfies the necessary conditions discussed earlier to achieve a stable wormhole solution. By utilizing this shape function, Eq.~(2.2) becomes
%\ie
%Z(r)= 
%\fe

%This embedding function \(Z(r)\) is depicted in Figure 1 where we have set \(r_{0}=1\). 

Here, the shape function $b(r)$ is considered for the scale factor $a(t) = a_{0} t^{n}$. 
In this case, the Kodama vector is observed to take the form
\ie
K_{ \pm}=\pm \frac{1}{2} a_0 n r t^{n - 1} - \frac{1}{2} \sqrt{\frac{r - r_0 \left(\frac{r_0}{r}\right)^{\frac{5 \lambda w + 3 \lambda + 3 w + 1}{\lambda w - \lambda + 3 w + 1}}}{r (\lambda + 1)}}
,
\fe
and, when evaluated on the trapping horizon, the Hawking temperature becomes

\begin{eqnarray}
T&=& \frac{t^{-n-2}}{8 \pi R_h^2 a_0 (\lambda + 1) (\lambda (w - 1) + 3w + 1)} \Bigg[  2 R_h^3 a_0^2 n t^{2n} (\lambda + 1) (n - 1) (\lambda (w - 1) + 3w + 1) \nonumber \\
&& + t^2 \Bigg( R_h \lambda (\lambda (w - 1) + 3w + 1) \quad - r_0 \left(\frac{r_0}{R_h}\right)^{\frac{\lambda (5w + 3) + 3w + 1}{\lambda (w - 1) + 3w + 1}} (\lambda (5w + 3) + 3w + 1) \nonumber \\
&& \quad - \Big( R_h \lambda + r_0 \left(\frac{r_0}{R_h}\right)^{\frac{\lambda (5w + 3) + 3w + 1}{\lambda (w - 1) + 3w + 1}} \Big) (\lambda (w - 1) + 3w + 1) \Bigg) \Bigg]. \label{Tdy}
\end{eqnarray}

\begin{figure}
    \centering
     \includegraphics[scale=0.5]{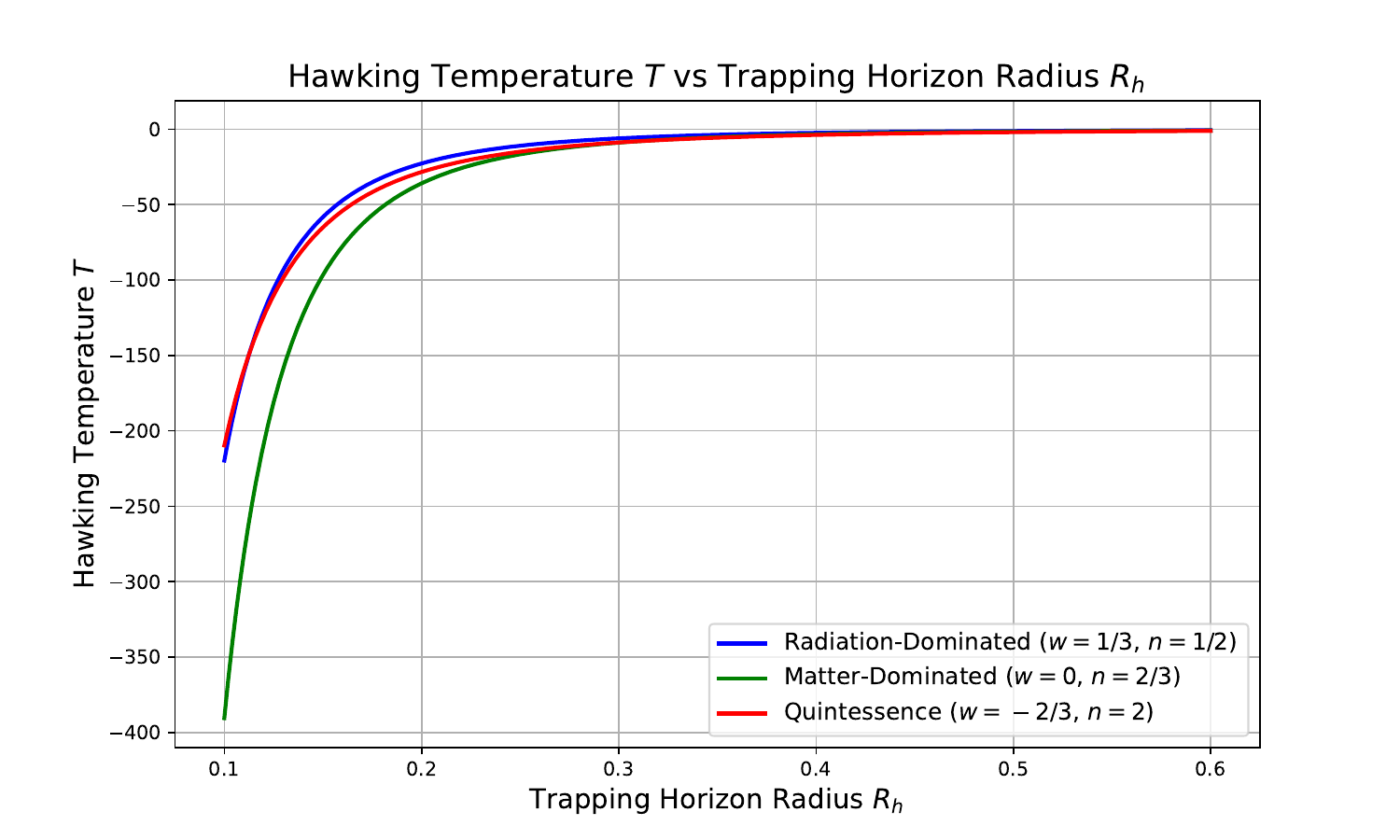}
    \caption{Hawking temperature $T$ as a function of the trapping horizon radius $R_h$ at a fixed time $t = 0.005$, for three cosmological scenarios: radiation--dominated ($w = 1/3$, $n = 1/2$), matter--dominated ($w = 0$, $n = 2/3$), and quintessence ($w = -2/3$, $n = 2$).}
    \label{Tdynamic}
\end{figure}

The Fig. \ref{Tdynamic} displays the Hawking temperature $T$ as a function of the trapping horizon radius $R_h$ shown in Eq. (\ref{Tdy}) for three cosmological scenarios. The temperature $T$ generally decreases as $R_h$ increases for all cases, highlighting an inverse relationship between the horizon size and temperature. The specific rate of decrease differs across the scenarios, with quintessence showing a steeper decline compared to the radiation-- and matter--dominated cases, reflecting the strong influence of $w$ and $n$  on the dynamics. This behavior aligns with theoretical expectations, as smaller horizons correspond to higher energy scales in the context of Hawking radiation.

Negative Hawking temperatures, as observed in the wormhole solutions, can have remarkable physical interpretations. Unlike black holes, which are thermodynamically associated with positive temperatures due to their event horizons emitting thermal radiation, wormholes lack such a traditional thermodynamic setup. Also, it often suggests non--standard thermal behavior, such as reversed heat flow or the system existing in a nonequilibrium thermodynamic state. These discussions are strictly conducted through geometric analysis. Conversely, the thermodynamic properties will be addressed using ensemble theory.

%%%%%%%%%%%%%%%%%%%%%%%%%%%%%%%%%%%%%%%%%%%%%%%%%%%%%%%%%%%%%%%%%%%%%%%%%%%%%%%%%%%%%%%%%%%%%%%%%%%%%%%%%%%%%%%%%%%%%%%%%%%%%%%%%%%%%%%%%%%%%%%%%%%%%%%%%%%%%%%%%%%%%%%%%%%%%%%%%%%%%%%%%%%%%%%%%%%%%%%%%%%%%%%%%%%%%%%%%%%%%%%%%%%%%%%%%%%%%%%%%%%%%%%%%%%%%%%%%%%%%%%%%%%%%%%%%%%%%%%%%%%%%%%%%%%%%%%%%%%%%%%%%%%%%%%%%%%%%%%%%%%%%%%%%%%%%%%%%%%%%%%%%%%%%%%%

\section{Thermodynamics via ensemble theory}

In this section, we analyze the thermodynamic properties of particles associated with de Broglie matter waves within the framework of a bumblebee wormhole. Our approach begins by introducing the following dispersion relation:
\ie
\label{energymomentumrelation1}
\hbar^{2}E^{2} = m^{2}_{0}c^{4} \Omega^{2}\left(1 + \frac{\Phi^2 \hbar^{2}{\bf{k}}^{2}}{(m_0 c)^{2}}\right). \nonumber
\fe

The exploration of thermodynamic properties under the framework of MDRs is essential for advancing our understanding of astrophysical phenomena \cite{amelino2013quantum}. By investigating their thermodynamic effects, we can more accurately detect and interpret unique features in observational data. This includes potential anomalies such as gamma--ray burst time delays \cite{jacob2008lorentz}, shifts in ultra--high energy cosmic ray spectra \cite{anchordoqui2003ultrahigh}, and deviations in the radiation patterns of active galactic nuclei \cite{anchordoqui2003ultrahigh}. Additionally, the authors have previously conducted a related analysis, applying \textit{rainbow} gravity \cite{aaa33} and investigating the thermodynamics of the Ellis wormhole \cite{araujo2023thermodynamical}.

As expected, the correlation between energy and momentum in Eq. (\ref{asddasd}) differs from that of photon--like particle modes. This variation leads to several noteworthy remarks, which will be discussed below. It is important to note that, from this point forward, we adopt natural units for our calculations. From the above expression, two distinct solutions emerge, but only one aligns with our goal of obtaining real, positive--definite values:
\begin{equation}
{\bf{k}} = \frac{\sqrt{-E^2-m_0^2}}{\sqrt{\frac{r-r_{0} \left(\frac{r_{0}}{r}\right)^{\frac{3 \lambda +(5 \lambda +3) w+1}{-\lambda +(\lambda +3) w+1}}}{(\lambda +1) r}}}
\end{equation}
so that we can address $\mathrm{d} {\bf{k}}$
\ie
\mathrm{d} {\bf{k}} =-\frac{E}{\sqrt{-E^2-m_0^2} \sqrt{\frac{r-r_{0} \left(\frac{r_{0}}{r}\right)^{\frac{3 \lambda +(5 \lambda +3) w+1}{-\lambda +(\lambda +3) w+1}}}{(\lambda +1) r}}}.
\label{vol}
\fe
At this stage, we proceed with the integration over momentum space to determine the accessible states of the system
\ie
\Xi(E) = \int_{0}^{\infty} {\bf{k}}^{2}\,\mathrm{d}{\bf{k}}.\label{ms2}
\fe
Here, \(\Gamma\) represents the volume of the thermal reservoir. In this context, Eq. (\ref{ms2}) becomes
\begin{widetext}
\ie
\begin{split}
\Xi(r,r_{0}) & =  -\int_{0}^{\infty} \frac{E \sqrt{-E^2-m_0^2}}{\left(\frac{r-r_{0} \left(\frac{r_{0}}{r}\right)^{\frac{3 \lambda +(5 \lambda +3) w+1}{-\lambda +(\lambda +3) w+1}}}{(\lambda +1) r}\right)^{3/2}} \,\mathrm{d}E.
\end{split}
\fe
\end{widetext}
For clarity, all calculations in this manuscript are presented on a ``per volume" basis. To enhance the reader's understanding, we first introduce the most general form of the partition function \cite{greiner2012thermodynamics}: 
\ie
Z = \frac{1}{N!h^{3N}} \int \mathrm{d}q^{3N}\mathrm{d}k^{3N} e^{-\beta H(q,p)}  \equiv \int \mathrm{d}E \,\Xi(E) e^{-\beta E}. \label{partti1}
\fe
The expression applies to an indistinguishable, spinless gas. Planck's constant is denoted by \( h \), the Boltzmann constant by \( \kappa_{B} \), and \( \beta = 1/\kappa_{B}T \). The generalized momenta and coordinates are represented by \( k \) and \( q \), respectively, while \( H(k,q) \) describes the system's Hamiltonian, and \( N \) refers to the number of particles. However, the spin of the particles is not considered in Eq. (\ref{partti1}). To remedy this, the spin contribution must be incorporated as follows \cite{isihara2013statistical,wannier1987statistical,salinas1999introduccao,vogt2017statistical,mandl1991statistical}:
\ie
\mathrm{ln}[Z] = \int \mathrm{d}E \,\Xi(E) \mathrm{ln} [ 1- e^{-\beta E}],
\fe
with \(\mathrm{ln} [ 1- e^{-\beta E}]\) corresponding to bosons, following the Bose--Einstein distribution. Consequently, the partition function can be expressed as:
\begin{widetext}
\ie\label{lnz1}
\begin{split}
\mathrm{ln}[Z(r,r_{0})] & = \frac{E \sqrt{-E^2-m_0^2} \, \ln \left(1-e^{-\beta  E}\right)}{\left(\frac{r-r_{0} \left(\frac{r_{0}}{r}\right)^{\frac{3 \lambda +(5 \lambda +3) w+1}{-\lambda +(\lambda +3) w+1}}}{(\lambda +1) r}\right)^{3/2}} \mathrm{d}E.
\end{split}
\fe
\end{widetext}
Based on the above expression, the relevant thermal quantities will be explored in detail in the upcoming sections. Although \textit{analytical} solutions are possible, a \textit{numerical} approach will be adopted for the analysis. The thermodynamic functions are defined as follows:
\ie
\begin{split}
  & P (r,r_{0})= \frac{1}{\beta} \mathrm{ln}\left[Z(r,r_{0})\right], \\
 & U(r,r_{0})=-\frac{\partial}{\partial\beta} \mathrm{ln}\left[Z(r,r_{0})\right], \\
 & S(r,r_{0})=k_B\beta^2\frac{\partial}{\partial\beta}F(r,r_{0}), \\
 & C(r,r_{0})=-k_B\beta^2\frac{\partial}{\partial\beta}U(r,r_{0}),
\label{properties}
\end{split}
\fe  
Here, \(P(r, r_{0})\) represents the pressure, \(U(r, r_{0})\) the mean energy, \(S(r, r_{0})\) the entropy, and \(C(r, r_{0})\) the heat capacity. In the next subsection, the equation of state will be examined. From this point forward, all thermal analyses will focus on massless particles, where \(m_{0} \to 0\).

%%%%%%%%%%%%%%%%%%%%%%%%%%%%%%%%%%%%%%%%%%%%%%%%%%%%%%%%%%%%%%%%%%%%%%%%%%%%%%%%%%%%%%%%%%%%%%%%%%%%%%%%%%%%%%%%%%%%%%%%%%%%%%%%%%%%%%%%%%%%%%%%%%%%%%%%%%%%%%%%%%%%%%%%%%%%%%%%%%%%%%%%%%%%%%%%%%%%%%%%%%%%%%%%%%%%%%%%%%%%%%%%%%%%%%%%%%%%%%%%%%%%%%%%%%%%%%%%%%%%%%%%%%%%%%%%%%%%%%%%%%%%%%%%%%%%%%%%%%%%%%%%%%%%%%%%%%%%%%%%%%%%%%%%%%%%%%%%%%%%%%%%%%%%%%%%

\subsection{Pressure}

\begin{figure}
    \centering
     \includegraphics[scale=0.5]{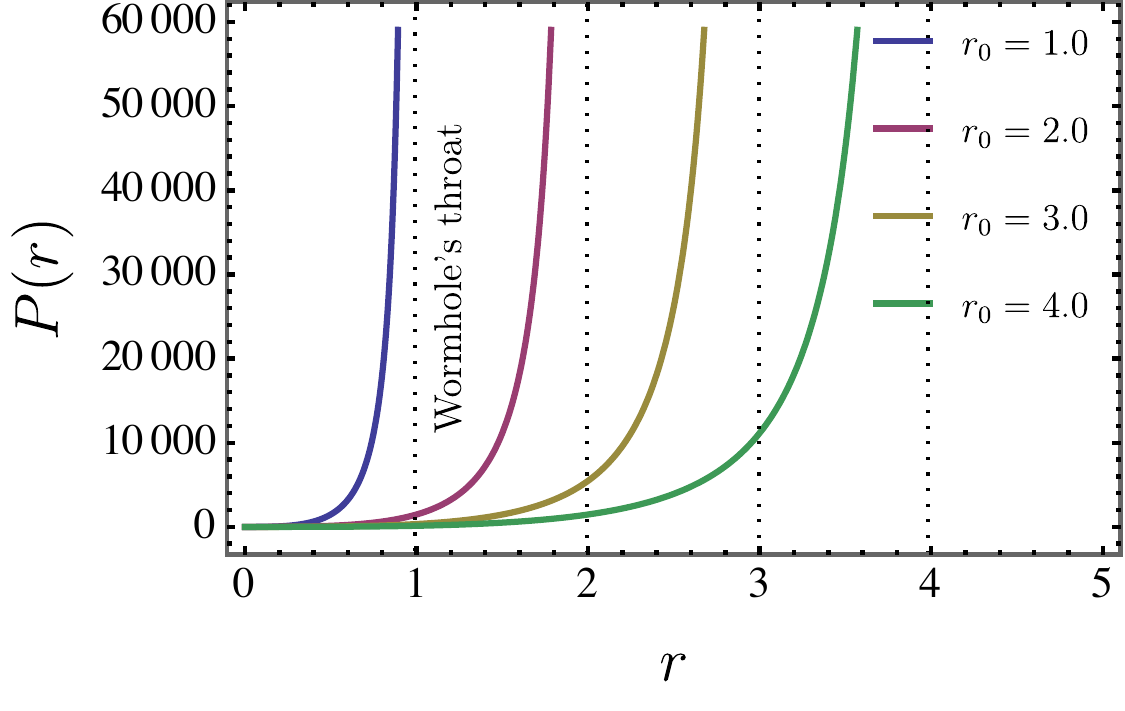}
    \caption{$P(r)$ as a function of $r$ for different values of the wormhole's throat $r_{0}$ and $T=10 K$.}
    \label{pronly}
\end{figure}

Initially, we write our first thermodynamic state quantity below (the pressure)
\ie
\begin{split}
\label{pressure}
P (r,r_{0}) & = \int^{\infty}_{0} \frac{E \sqrt{-E^2} \ln \left(1-e^{-\beta  E}\right)}{\beta  \left(\frac{-r+r_{0} \left(\frac{r_{0}}{r}\right)^{\frac{3 \lambda +(5 \lambda +3) w+1}{-\lambda +(\lambda +3) w+1}}}{(\lambda +1) r}\right)^{3/2}} \mathrm{d}E \\
& = \frac{\pi ^4 T^{4}}{45 \left(\frac{-r+r_{0} \left(\frac{r_{0}}{r}\right)^{\frac{3 \lambda +(5 \lambda +3) w+1}{-\lambda +(\lambda +3) w+1}}}{(\lambda +1) r}\right)^{3/2}}.
\end{split}
\fe
We proceed by analyzing the behavior of \(P(r,r_{0})\) as a function of \(r\) for different values of \(r_{0}\). In Fig. \ref{pronly}, the pressure is depicted as a function of the radial coordinate \(r\), for several values of the wormhole throat radius \(r_{0}\), under a fixed temperature of \(T = 10\, \text{K}\). The plot reveals an asymptotic trend as \(r\) approaches \(r_{0}\), which is emphasized by the dashed lines.
In what follows, we present a detailed examination of the thermodynamic properties of the system as they vary with temperature. To achieve this, we will investigate three separate scenarios, each offering insights into the behavior of the corresponding equations of state, as described in the sections below.

%%%%%%%%%%%%%%%%%%%%%%%%%%%%%%%%%%%%%%%%%%%%%%%%%%%%%%%%%%%%%%%%%%%%%%%%%%%%%%%%%%%%%%%%%%%%%%%%%%%%%%%%%%%%%%%%%%%%%%%%%%%%%%%%%%%%%%%%%%%%%

\subsubsection{Asymptotically far}

\begin{figure}
    \centering
     \includegraphics[scale=0.5]{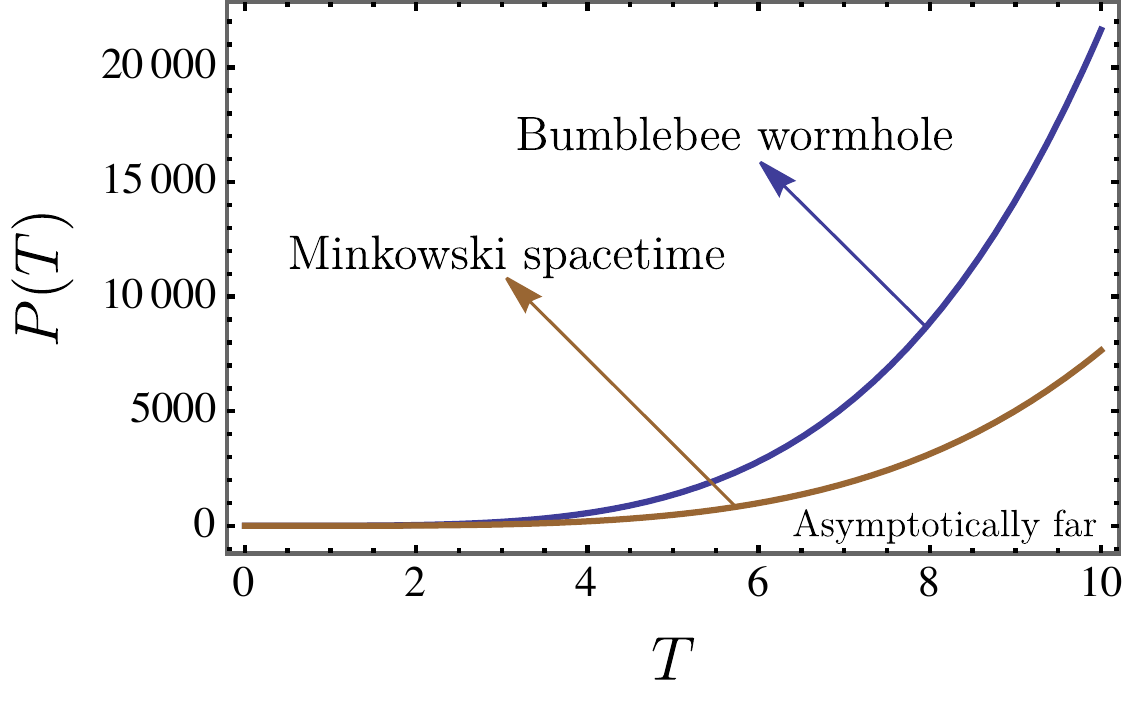}
    \caption{The comparison of the pressure of photon--like particles in Minkowski and bumblebee wormhole spacetimes.}
    \label{MB}
\end{figure}

At this point, we consider the limit \(\lim\limits_{r \to \infty}P(r,r_{0})\), which allows us to determine the corresponding value
\ie
\lim\limits_{r \to \infty}P(r,r_{0}) = P (T) =\frac{\pi ^4 T^{4}}{90 \sqrt{2} }.
\fe
It is important to highlight that this result shows a slight deviation from the expression obtained for photons in Minkowski spacetime, where \(P(T) = \pi^4 T^4/45\). This discrepancy is primarily due to the non--asymptotically flat nature of the bumblebee wormhole spacetime \cite{ovgun2019exact}. To facilitate a more direct comparison between our findings and the results in Minkowski spacetime, we provide Fig. \ref{MB}.

%%%%%%%%%%%%%%%%%%%%%%%%%%%%%%%%%%%%%%%%%%%%%%%%%%%%%%%%%%%%%%%%%%%%%%%%%%%%%%%%%%%%%%%%%%%%%%%%%%%%%%%%%%%%%%%%%%%%%%%%%%%%%%%%%%%%%%%%%%%%%

\subsubsection{Close to the throat}

\begin{figure}
    \centering
     \includegraphics[scale=0.5]{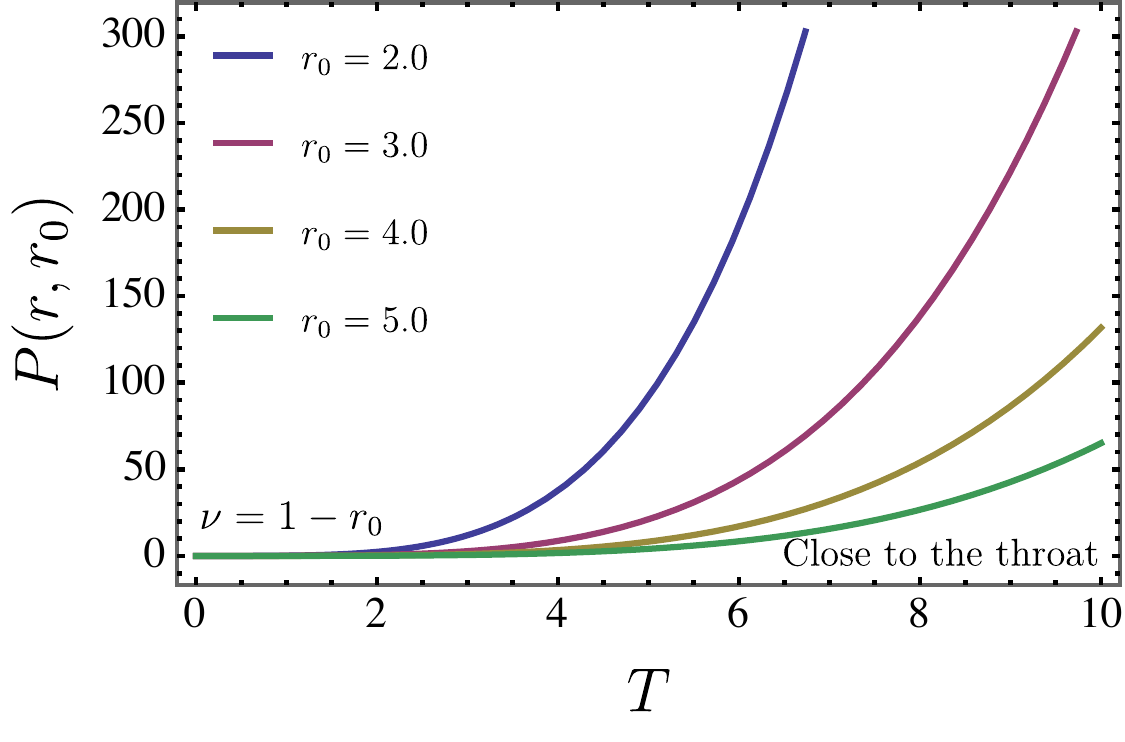}
    \caption{The pressure $P(r,r_{0})$ as a function of temperature $T$ for different values of the wormhole`s throat $r_{0}$, when we consider a configuration located close to it.}
    \label{close}
\end{figure}

In this subsection, we analyze the thermodynamic characteristics of the system in the vicinity of the wormhole throat. To facilitate this, we redefine the radial coordinate as \(r = r_{0} + \nu\), where \(\nu\) denotes a small offset from the throat. In particular, we choose \(\nu = 1 - r_{0}\) to represent this configuration. With these assumptions, the pressure can be expressed as follows:
\ie
P (r,r_{0}) = \frac{\pi^4 T^4}{45 \left(\frac{r_{0} \left(\left(\frac{r_{0}}{\nu +r_{0}}\right)^{\frac{3 \lambda +(5 \lambda +3) w+1}{-\lambda +(\lambda +3) w+1}}-1\right)-\nu }{(\lambda +1) (\nu + r_{0})}\right)^{3/2}}.
\fe
We now turn our attention to the behavior of this quantity as illustrated in Fig. \ref{close}. The plots corresponding to various \(r_{0}\) values exhibit a tendency to converge. Nevertheless, as will be demonstrated in the following subsection, this pattern significantly differs from the behavior observed near the wormhole’s throat.

%%%%%%%%%%%%%%%%%%%%%%%%%%%%%%%%%%%%%%%%%%%%%%%%%%%%%%%%%%%%%%%%%%%%%%%%%%%%%%%%%%%%%%%%%%%%%%%%%%%%%%%%%%%%%%%%%%%%%%%%%%%%%%%%%%%%%%%%%%%%%

\subsubsection{Very close to the throat}

We now focus on the behavior in the immediate vicinity of the bumblebee wormhole’s throat. A pertinent question arises: why not carry out the calculations directly at the throat? The answer lies in the fact that doing so leads to divergences in the results. For this reason, we instead perform the analysis near \(r_{0}\).

\begin{figure}
    \centering
     \includegraphics[scale=0.425]{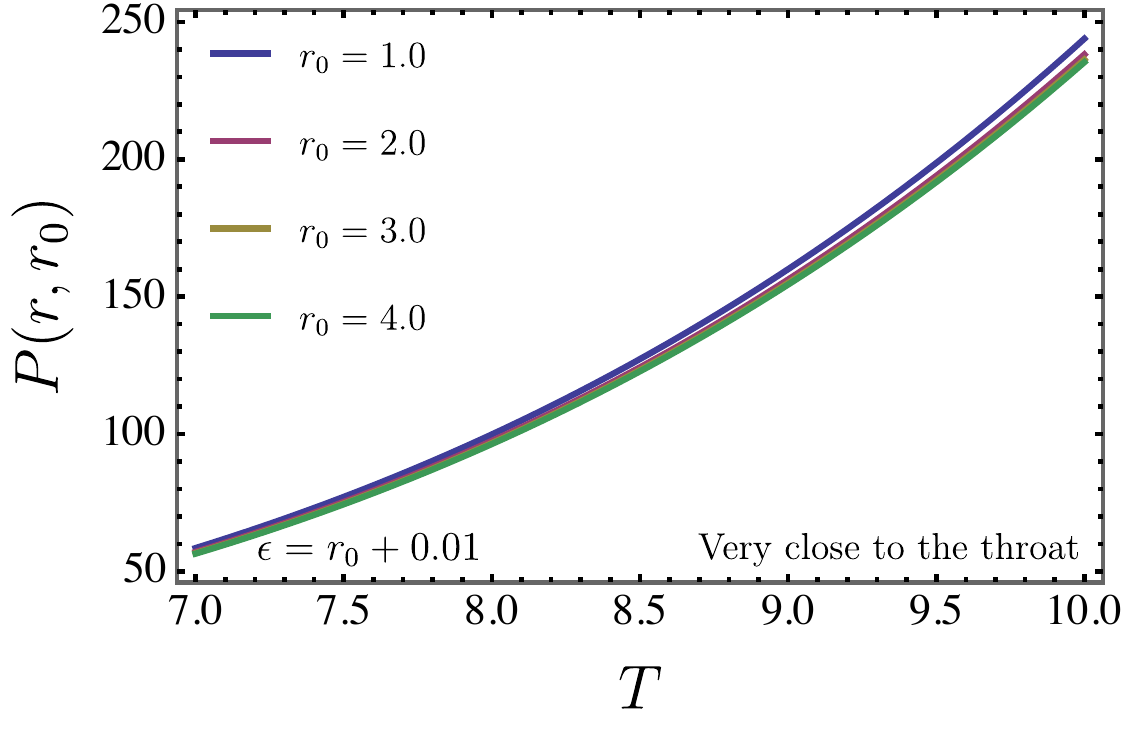}
     \includegraphics[scale=0.425]{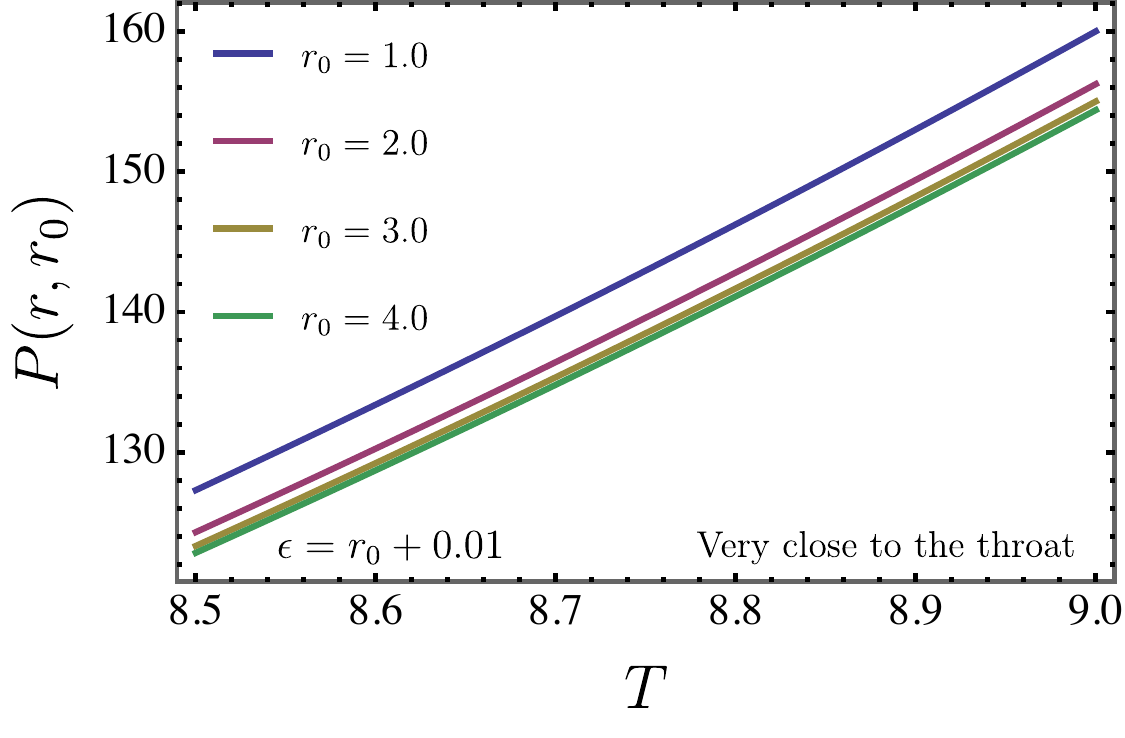}
    \caption{The pressure $P(r,r_{0})$ as a function of temperature $T$ for different values of the wormholes throat $r_{0}$, when we consider a configuration located very close to it.}
    \label{veryclose}
\end{figure}

To proceed, we define \(r = \epsilon + r_{0}\), where \(\epsilon\) is a small perturbation parameter. Expanding Eq. (\ref{pressure}) under this assumption yields the following expression:
\ie
P (r,r_{0}) = \frac{\pi ^4 \left(\frac{1}{r_{0}}\right)^{3/2} T^4 \epsilon ^{3/2}}{3072}+\frac{\pi ^4 T^4}{360 \left(\frac{1}{r_{0}}\right)^{3/2} \epsilon ^{3/2}}+\frac{13 \pi ^4 \sqrt{\frac{1}{r_{0}}} T^4 \sqrt{\epsilon }}{3840}-\frac{\pi ^4 T^4}{160 \sqrt{\frac{1}{r_{0}}} \sqrt{\epsilon }}.
\fe
By expanding the expression up to second order in \(\epsilon\), we obtain the relevant terms. To provide a more intuitive understanding of this result, we present Fig. \ref{veryclose}, which shows the pressure calculated near \(r_{0}\), where \(\epsilon = r_{0} + 0.01\), across different values of \(r_{0}\). Unlike the earlier scenario, where behavior near the throat was studied, this plot reveals a clear divergence for small \(\epsilon\), as anticipated.

%%%%%%%%%%%%%%%%%%%%%%%%%%%%%%%%%%%%%%%%%%%%%%%%%%%%%%%%%%%%%%%%%%%%%%%%%%%%%%%%%%%%%%%%%%%%%%%%%%%%%%%%%%%%%%%%%%%%%%%%%%%%%%%%%%%%%%%%%%%%%%%%%%%%%%%%%%%%%%%%%%%%%%%%%%%%%%%%%%%%%%%%%%%%%%%%%%%%%%%%%%%%%%%%%%%%%%%%%%%%%%%%%%%%%%%%%%%%%%%%%%%%%%%%%%%%%%%%%%%%%%%%%%%%%%%%%%%%%%%%%%%%%%%%%%%%%%%%%%%%%%%%%%%%%%%%%%%%%%%%%%%%%%%%%%%%%%%%%%%%%%%%%%%%%%%%

\subsection{Mean energy}

\begin{figure}
    \centering
     \includegraphics[scale=0.5]{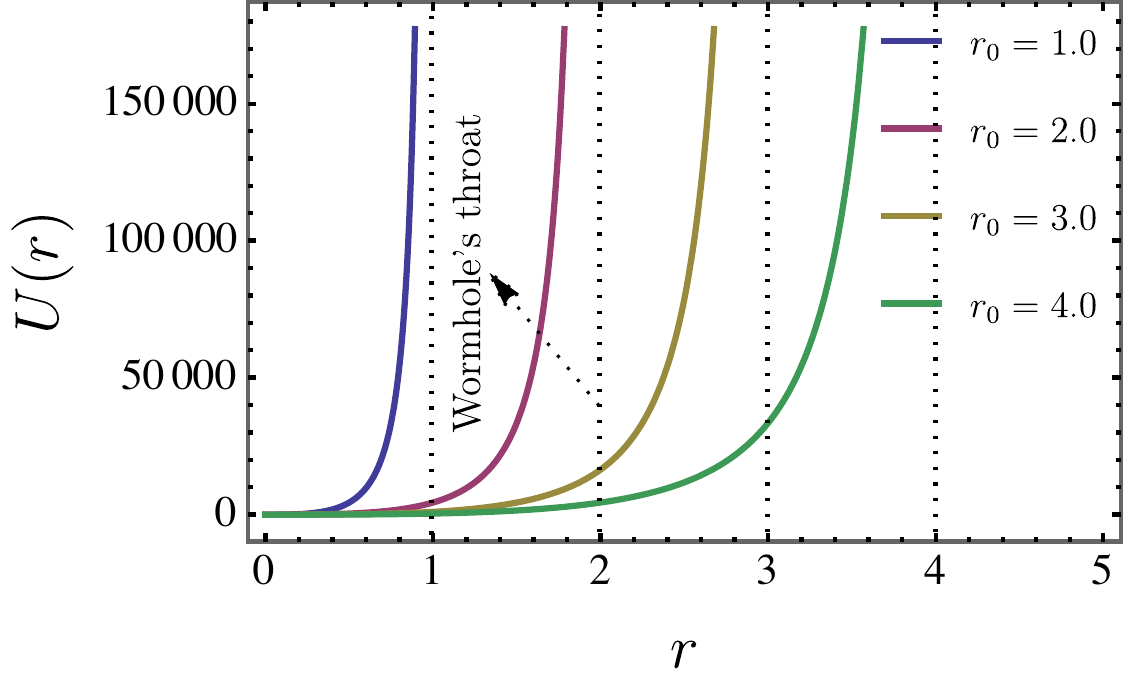}
    \caption{The mean $U(r)$ as a function of $r$ for different configurations of $r_{0}$ and $T=10 K$.}
    \label{ur}
\end{figure}

We now turn our attention to the general expression for the mean energy
\ie
\begin{split}
\label{pressureU}
U (r,r_{0}) & = -\int^{\infty}_{0} \frac{E^2 \sqrt{-E^2} e^{-\beta  E}}{\left(1-e^{-\beta  E}\right) \left(\frac{r-r_{0}\left(\frac{r_{0}}{r}\right)^{\frac{3 \lambda +(5 \lambda +3) w+1}{-\lambda +(\lambda +3) w+1}}}{(\lambda +1) r}\right)^{3/2}} \mathrm{d}E \\
& = \frac{\pi ^4 T^4}{15 \left(\frac{r_{0} \left(\frac{r_{0}}{r}\right)^{\frac{3 \lambda +(5 \lambda +3) w+1}{-\lambda +(\lambda +3) w+1}}-r}{(\lambda +1) r}\right)^{3/2}}.
\end{split}
\fe
We focus here on the behavior of \(U(r,r_{0})\) at a constant temperature of \(10\,K\), as illustrated in Fig. \ref{ur}. Much like the pressure \(P(r,r_{0})\), the mean energy \(U(r,r_{0})\) exhibits a divergence at the throat, \(r_{0}\). In what follows, we analyze three specific cases to investigate the corresponding equations of state, outlined as follows.

%%%%%%%%%%%%%%%%%%%%%%%%%%%%%%%%%%%%%%%%%%%%%%%%%%%%%%%%%%%%%%%%%%%%%%%%%%%%%%%%%%%%%%%%%%%%%%%%%%%%%%%%%%%%%%%%%%%%%%%%%%%%%%%%%%%%%%%%%%%%%

\subsubsection{Asymptotically far}

\begin{figure}
    \centering
     \includegraphics[scale=0.5]{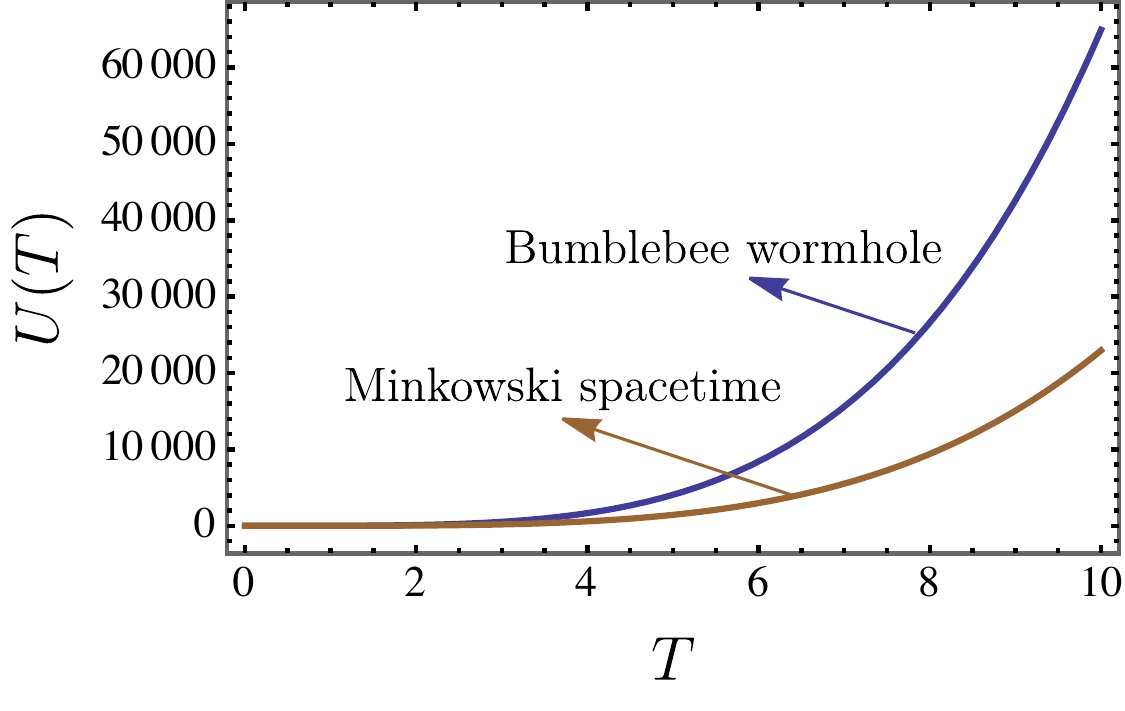}
    \caption{The comparison of the mean energy of photon--like particles in Minkowski and bumblebee wormhole spacetimes.}
    \label{MBU}
\end{figure}

Next, we evaluate the limit \(\lim\limits_{r \to \infty} P(r, r_{0})\), resulting in the corresponding value:
\ie
\lim\limits_{r \to \infty}U(r,r_{0}) = U (T) =\frac{\pi ^4 T^{4}}{30 \sqrt{2} }.
\fe
It is essential to point out that this result deviates slightly from the one derived for photons in Minkowski spacetime, where \(U(T) = \frac{\pi^4 T^4}{15}\). This difference stems, in part, from the fact that the bumblebee wormhole resides in a non--asymptotically flat spacetime \cite{ovgun2019exact}. To better illustrate the contrast between our findings and those in Minkowski spacetime, we include Fig. \ref{MBU}.

%%%%%%%%%%%%%%%%%%%%%%%%%%%%%%%%%%%%%%%%%%%%%%%%%%%%%%%%%%%%%%%%%%%%%%%%%%%%%%%%%%%%%%%%%%%%%%%%%%%%%%%%%%%%%%%%%%%%%%%%%%%%%%%%%%%%%%%%%%%%%

\subsubsection{Close to the throat}

\begin{figure}
    \centering
     \includegraphics[scale=0.5]{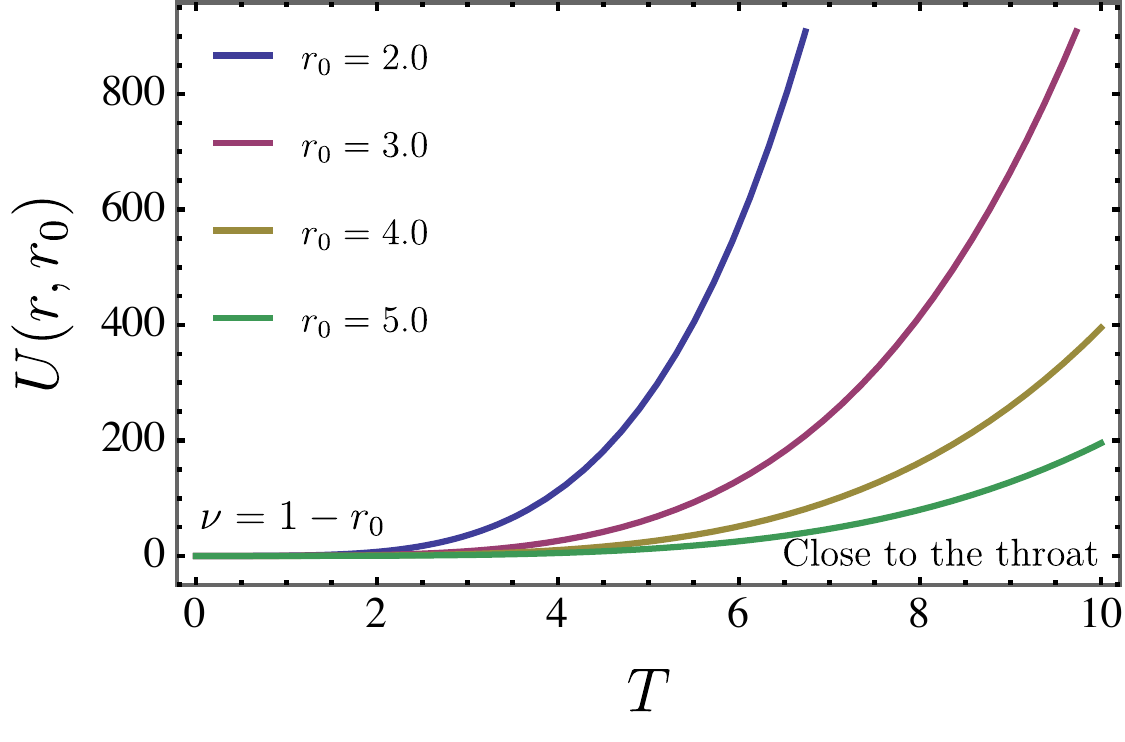}
    \caption{The mean energy $U(r,r_{0})$ as a function of temperature $T$ for different values of the wormhole`s throat $r_{0}$, when we consider a configuration located close to it.}
    \label{closeU}
\end{figure}

In this subsection, we investigate the thermodynamic behavior of the system in the vicinity of the throat. For this analysis, we define \(r = r_{0} - \nu\), where \(\nu\) represents a small deviation from the throat. In particular, we choose \(\nu = 1 - r_{0}\) to capture this configuration. Under these assumptions, the mean value is expressed as:
\ie
U (r,r_{0}) = \frac{\pi ^4 T^4}{15 \left(\frac{\nu -r_{0} \left(\frac{r_{0}}{\nu +r_{0}}\right)^{\frac{3 \lambda +(5 \lambda +3) w+1}{-\lambda +(\lambda +3) w+1}}+r_{0}}{(\lambda +1) (\nu +r_{0})}\right)^{3/2}}.
\fe
We turn to Fig. \ref{closeU} to analyze the behavior of this quantity. The plots for various values of \(r_{0}\) show a tendency to converge. However, as we will explore in the following subsection, this behavior contrasts with what occurs when the system is positioned much closer to the wormhole's throat.

%%%%%%%%%%%%%%%%%%%%%%%%%%%%%%%%%%%%%%%%%%%%%%%%%%%%%%%%%%%%%%%%%%%%%%%%%%%%%%%%%%%%%%%%%%%%%%%%%%%%%%%%%%%%%%%%%%%%%%%%%%%%%%%%%%%%%%%%%%%%%

\subsubsection{Very close to the throat}

We now focus on the behavior in the immediate vicinity of the bumblebee wormhole’s throat. A reasonable question to ask is why not perform the calculations exactly at the throat. The reason is that doing so leads to divergences in the results. Consequently, we carry out the analysis near \(r_{0}\) instead.

\begin{figure}
    \centering
     \includegraphics[scale=0.425]{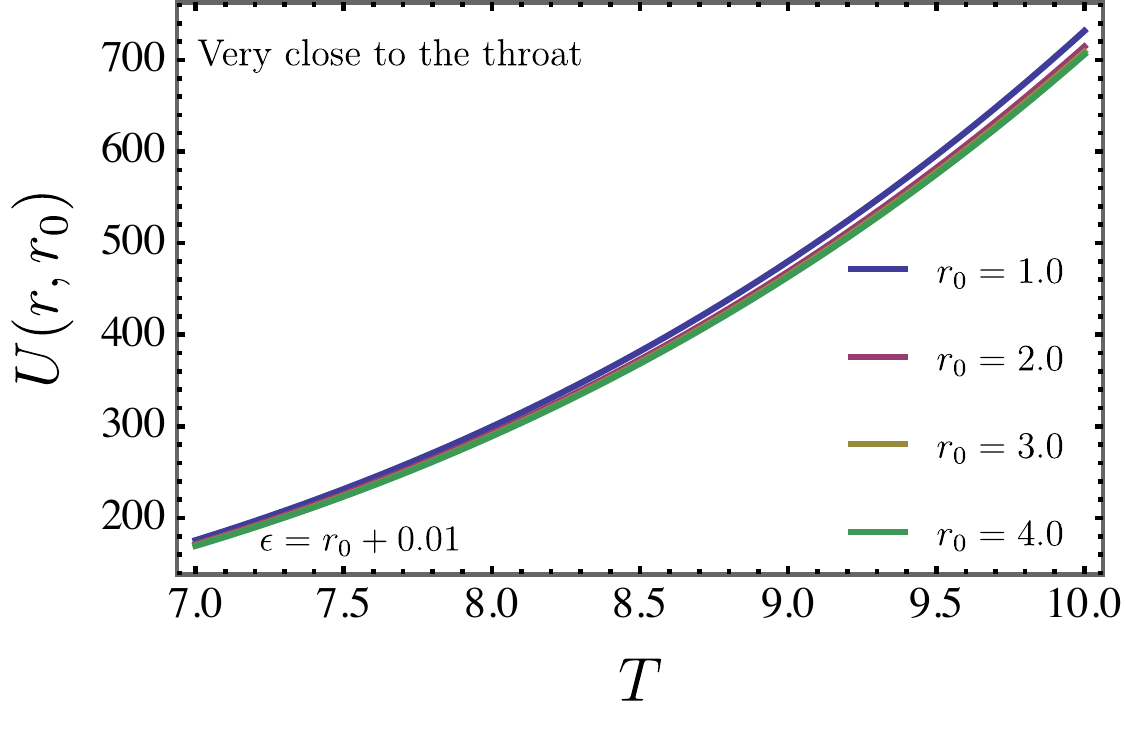}
     \includegraphics[scale=0.425]{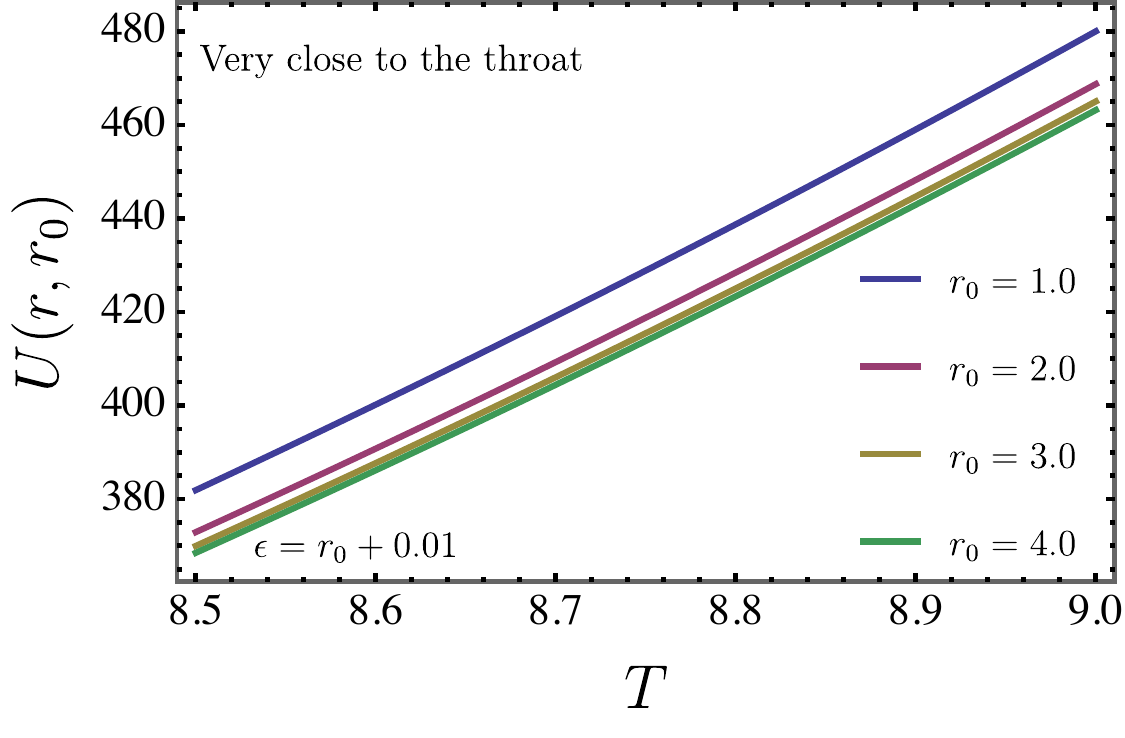}
    \caption{The mean energy $U(r,r_{0})$ as a function of temperature $T$ for different values of the wormhole`s throat $r_{0}$, when we consider a configuration located very close to it.}
    \label{verycloseU}
\end{figure}

For this purpose, we define \(r = \epsilon + r_{0}\), where \(\epsilon\) is a small parameter. Expanding Eq. (\ref{pressureU}) under these conditions gives the following expression:
\ie
U (r,r_{0}) = \frac{\pi ^4 \left(\frac{1}{r_{0}}\right)^{3/2} T^4 \epsilon ^{3/2}}{1024}+\frac{\pi ^4 T^4}{120 \left(\frac{1}{r_{0}}\right)^{3/2} \epsilon ^{3/2}}+\frac{13 \pi ^4 \sqrt{\frac{1}{r_{0}}} T^4 \sqrt{\epsilon }}{1280}-\frac{3 \left(\pi ^4 T^4\right)}{160 \sqrt{\frac{1}{r_{0}}} \sqrt{\epsilon }}.
\fe
Expanding up to second order in the parameter \(\epsilon\), we obtain the corresponding terms. To provide a clearer understanding of this result, Fig. \ref{verycloseU} illustrates the mean calculated near \(r_{0}\), with \(\epsilon = r_{0} + 0.01\) for different values of \(r_{0}\). Unlike the earlier case analyzed close to the throat, this plot reveals a clear divergence for small \(\epsilon\), as anticipated.

%%%%%%%%%%%%%%%%%%%%%%%%%%%%%%%%%%%%%%%%%%%%%%%%%%%%%%%%%%%%%%%%%%%%%%%%%%%%%%%%%%%%%%%%%%%%%%%%%%%%%%%%%%%%%%%%%%%%%%%%%%%%%%%%%%%%%%%%%%%%%%%%%%%%%%%%%%%%%%%%%%%%%%%%%%%%%%%%%%%%%%%%%%%%%%%%%%%%%%%%%%%%%%%%%%%%%%%%%%%%%%%%%%%%%%%%%%%%%%%%%%%%%%%%%%%%%%%%%%%%%%%%%%%%%%%%%%%%%%%%%%%%%%%%%%%%%%%%%%%%%%%%%%%%%%%%%%%%%%%%%%%%%%%%%%%%%%%%%%%%%%%%%%%%%%%%

\subsection{Entropy}

At this juncture, we direct our attention to the analysis of entropy
\ie
\begin{split}
\label{SSS}
S (r,r_{0}) & = -\int^{\infty}_{0} \frac{E \sqrt{-E^2} \left(\beta  E-\left(e^{\beta  E}-1\right) \ln \left(1-e^{-\beta  E}\right)\right)}{\left(e^{\beta  E}-1\right) \left(\frac{r-r_{0} \left(\frac{r_{0}}{r}\right)^{\frac{3 \lambda +(5 \lambda +3) w+1}{-\lambda +(\lambda +3) w+1}}}{(\lambda +1) r}\right)^{3/2}} \\
& = \frac{4 \pi ^4 T^3}{45 \left(\frac{r_{0} \left(\frac{r_{0}}{r}\right)^{\frac{3 \lambda +(5 \lambda +3) w+1}{-\lambda +(\lambda +3) w+1}}-r}{(\lambda +1) r}\right)^{3/2}}.
\end{split}
\fe

\begin{figure}
    \centering
     \includegraphics[scale=0.5]{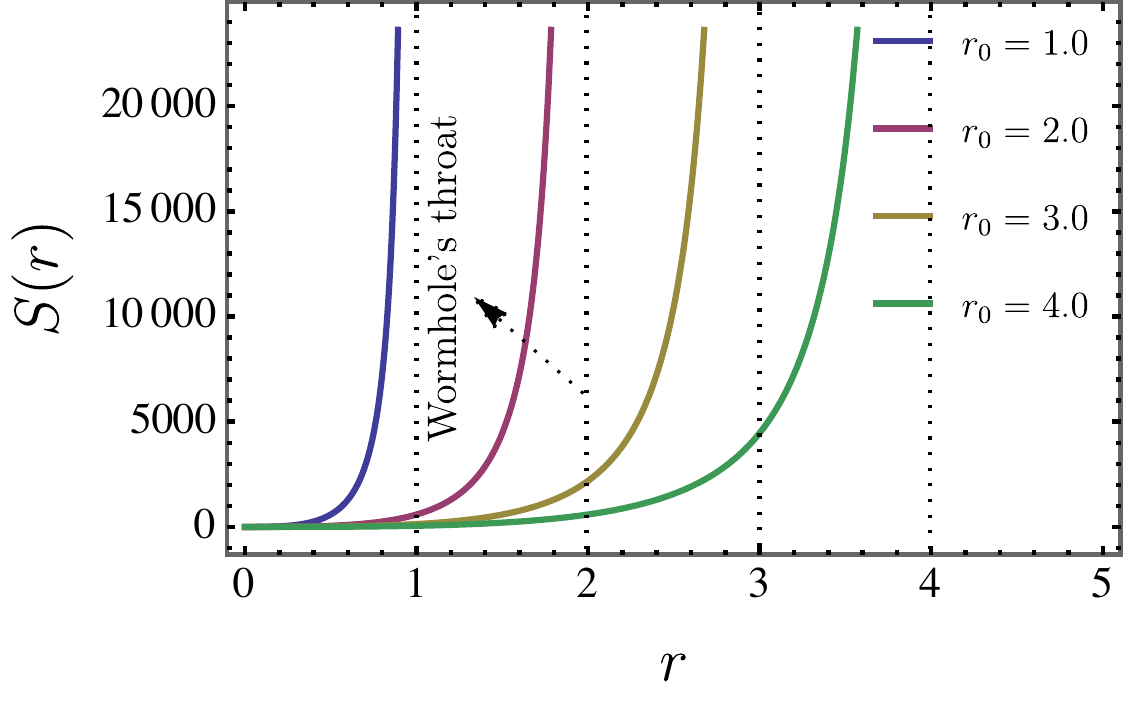}
    \caption{The entropy $S(r)$ as a function of $r$ for different configurations of $r_{0}$ and $T=10 K$.}
    \label{sr}
\end{figure}

The behavior of the equation is illustrated in Fig. \ref{sr}, where we observe that, much like the pressure and the mean energy, the entropy diverges at \(r_{0}\). At this point, we proceed by exploring three different cases to further analyze the behavior of the corresponding equations of state, as detailed below.

%%%%%%%%%%%%%%%%%%%%%%%%%%%%%%%%%%%%%%%%%%%%%%%%%%%%%%%%%%%%%%%%%%%%%%%%%%%%%%%%%%%%%%%%%%%%%%%%%%%%%%%%%%%%%%%%%%%%%%%%%%%%%%%%%%%%%%%%%%%%%

\subsubsection{Asymptotically far}

\begin{figure}
    \centering
     \includegraphics[scale=0.5]{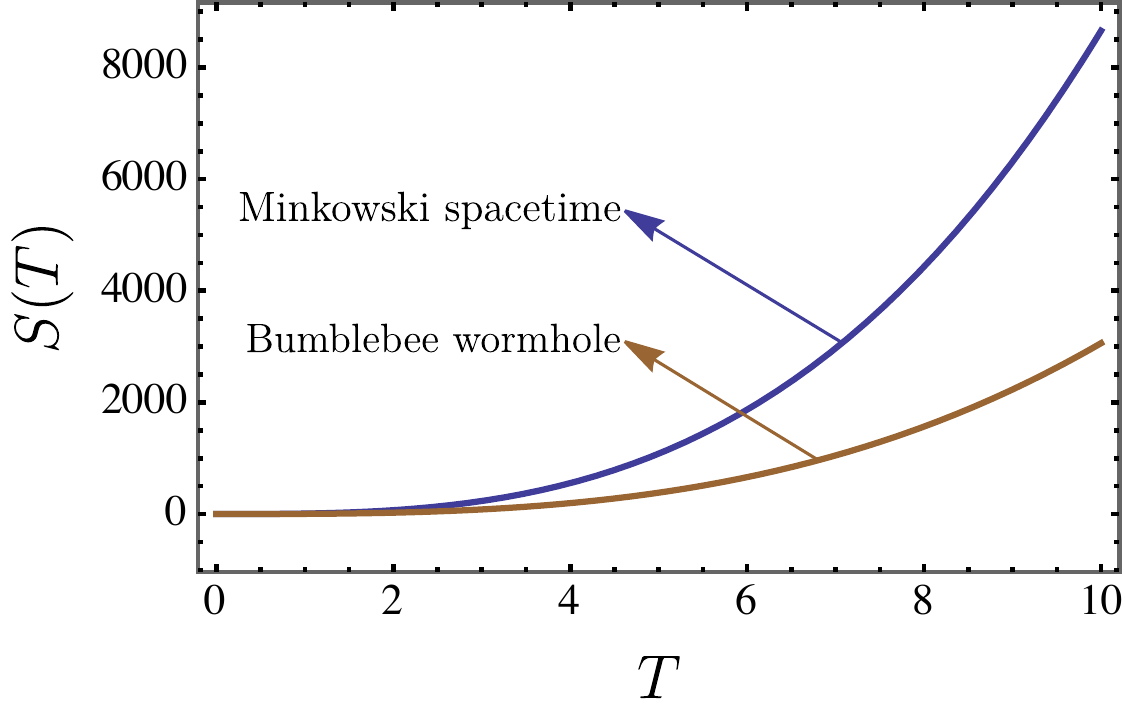}
    \caption{The comparison of the entropy of photon--like particles in Minkowski and bumblebee wormhole spacetimes.}
    \label{MBS}
\end{figure}

We now evaluate the limit \(\lim\limits_{r \to \infty} S(r, r_{0})\), which results in the corresponding value:
\ie
\lim\limits_{r \to \infty}S(r,r_{0}) = S (T) = \frac{1}{45} \sqrt{2} \pi ^4 T^3.
\fe
It is worth mentioning that this result shows a slight deviation from the one found for photons in Minkowski spacetime, where \(S(T) = \frac{4 \pi^4 T^3}{45}\). This difference stems partly from the non--asymptotically flat nature of the bumblebee wormhole spacetime \cite{ovgun2019exact}. For a more direct comparison between our findings and those in Minkowski spacetime, Fig. \ref{MBS} is provided.

%%%%%%%%%%%%%%%%%%%%%%%%%%%%%%%%%%%%%%%%%%%%%%%%%%%%%%%%%%%%%%%%%%%%%%%%%%%%%%%%%%%%%%%%%%%%%%%%%%%%%%%%%%%%%%%%%%%%%%%%%%%%%%%%%%%%%%%%%%%%%

\subsubsection{Close to the throat}

\begin{figure}
    \centering
     \includegraphics[scale=0.5]{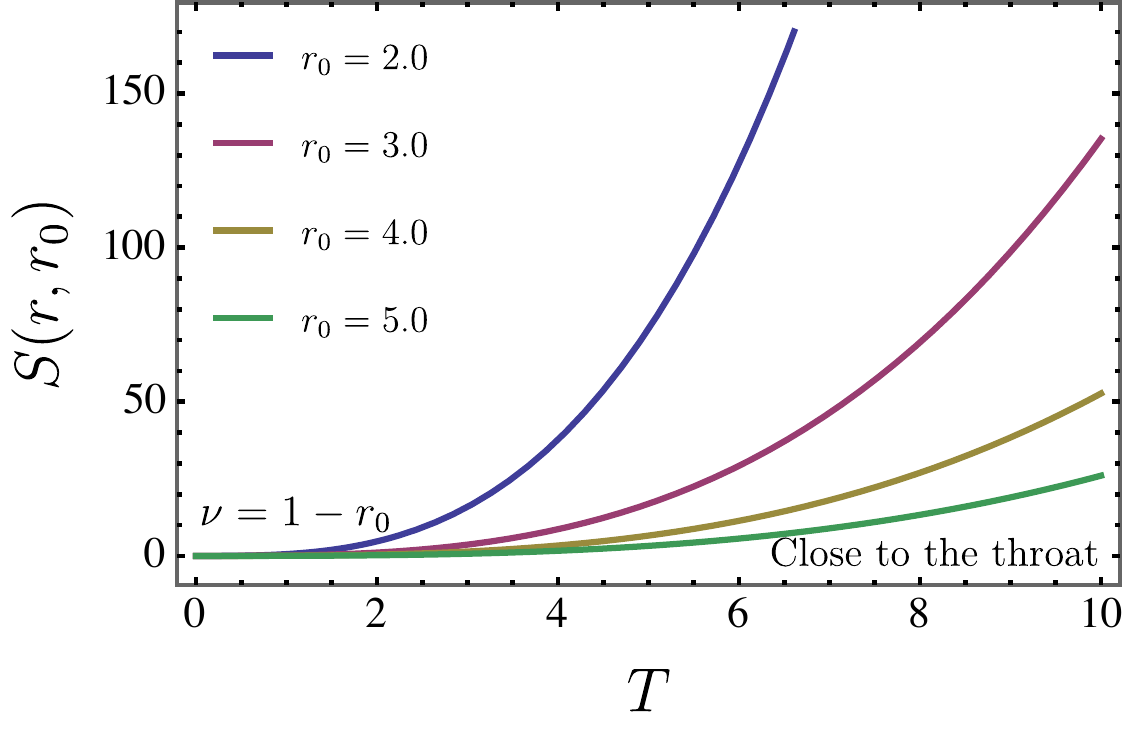}
    \caption{The entropy $S(r,r_{0})$ as a function of temperature $T$ for different values of the wormhole`s throat $r_{0}$, when we consider a configuration located close to it.}
    \label{closeS}
\end{figure}

In this section, we focus on the thermodynamic behavior of the system in the vicinity of the throat. To investigate this, we define \(r = r_{0} - \nu\), where \(\nu\) represents a slight deviation from the throat. Specifically, we take \(\nu = 1 - r_{0}\). With these parameters, the entropy is given by:
\ie
S (r,r_{0}) = \frac{4 \pi ^4 T^3}{45 \left(\frac{-\nu +r_{0} \left(\frac{r_{0}}{\nu +r_{0}}\right)^{\frac{3 \lambda +(5 \lambda +3) w+1}{-\lambda +(\lambda +3) w+1}}-r_{0}}{(\lambda +1) (\nu +r_{0})}\right)^{3/2}}.
\fe
We now turn to Fig. \ref{closeS} to investigate the behavior of this quantity. The plots for various values of \(r_{0}\) show a tendency to converge. However, as we will explore in the next subsection, this behavior stands in stark contrast to the configurations located much closer to the wormhole’s throat.

%%%%%%%%%%%%%%%%%%%%%%%%%%%%%%%%%%%%%%%%%%%%%%%%%%%%%%%%%%%%%%%%%%%%%%%%%%%%%%%%%%%%%%%%%%%%%%%%%%%%%%%%%%%%%%%%%%%%%%%%%%%%%%%%%%%%%%%%%%%%%

\subsubsection{Very close to the throat}

\begin{figure}
    \centering
     \includegraphics[scale=0.425]{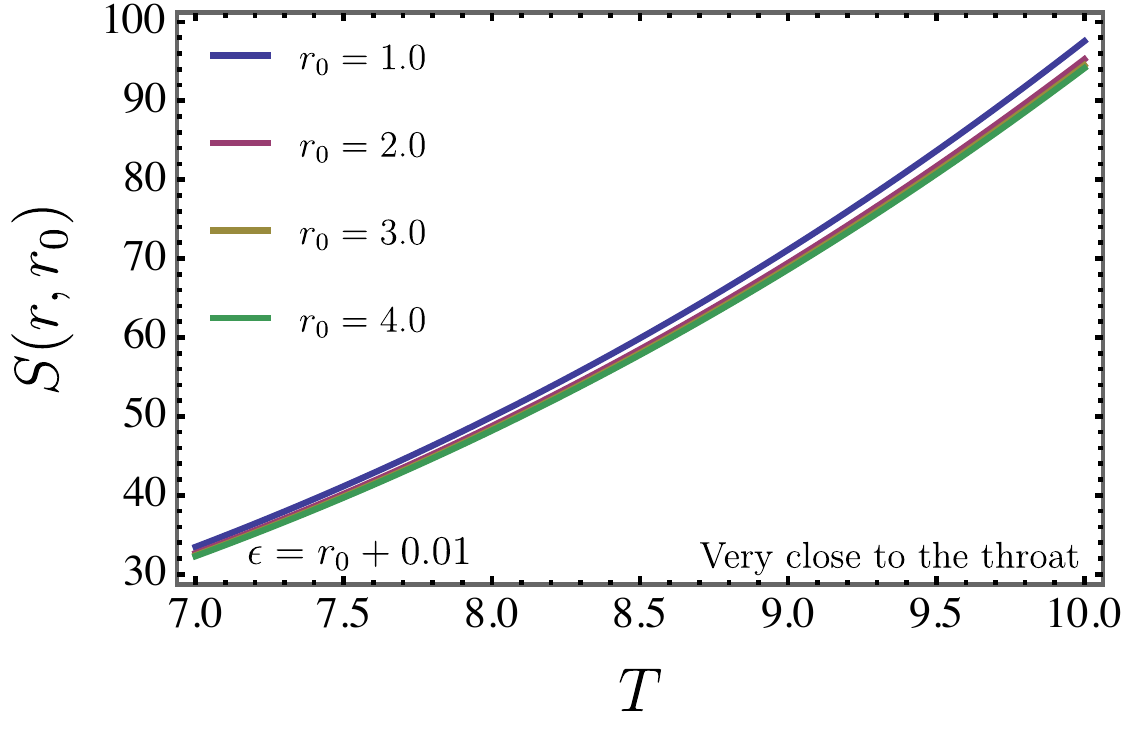}
     \includegraphics[scale=0.42]{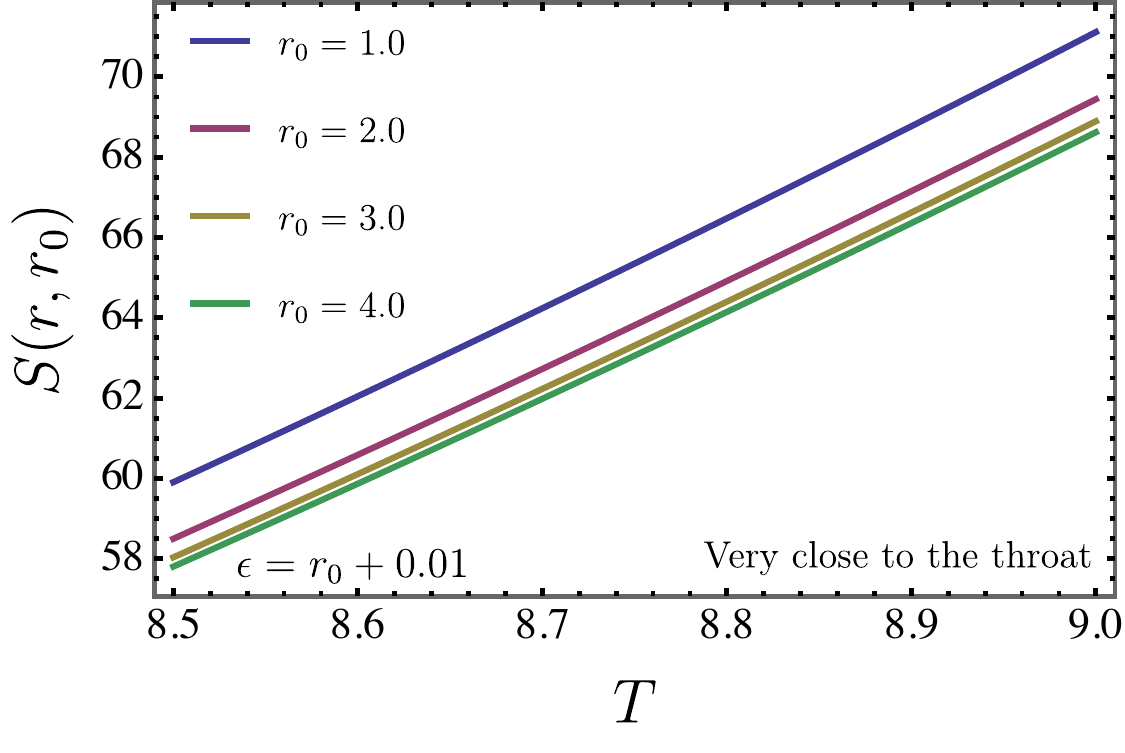}
    \caption{The entropy $S(r,r_{0})$ as a function of temperature $T$ for different values of the wormhole's throat $r_{0}$, when we consider a configuration located very close to it.}
    \label{verycloseS}
\end{figure}

We now focus on investigating the behavior near the throat of the bumblebee wormhole. One might ask why the calculations are not performed directly at the throat. The reason is that doing so leads to divergences in the results for this particular case. Instead, we carry out the calculations close to \(r_{0}\). To this end, we define \(r = \epsilon + r_{0}\), where \(\epsilon\) is a small parameter. By expanding Eq. (\ref{SSS}) under these conditions, we obtain the following expression:
\ie
S (r,r_{0}) = \frac{1}{768} \pi ^4 \left(\frac{1}{r_{0}}\right)^{3/2} T^3 \epsilon ^{3/2}+\frac{\pi ^4 T^3}{90 \left(\frac{1}{r_{0}}\right)^{3/2} \epsilon ^{3/2}}+\frac{13}{960} \pi ^4 \sqrt{\frac{1}{r_{0}}} T^3 \sqrt{\epsilon }-\frac{\pi ^4 T^3}{40 \sqrt{\frac{1}{r_{0}}} \sqrt{\epsilon }}.
\fe
Expanding the equation to second order in \(\epsilon\), we derive the necessary terms. For a clearer understanding, Fig. \ref{verycloseS} illustrates the entropy calculated near \(r_{0}\), with \(\epsilon = r_{0} + 0.01\) for various values of \(r_{0}\). In contrast to the previous analysis close to the throat, this plot reveals a clear divergence for small \(\epsilon\), as anticipated.

%%%%%%%%%%%%%%%%%%%%%%%%%%%%%%%%%%%%%%%%%%%%%%%%%%%%%%%%%%%%%%%%%%%%%%%%%%%%%%%%%%%%%%%%%%%%%%%%%%%%%%%%%%%%%%%%%%%%%%%%%%%%%%%%%%%%%%%%%%%%%%%%%%%%%%%%%%%%%%%%%%%%%%%%%%%%%%%%%%%%%%%%%%%%%%%%%%%%%%%%%%%%%%%%%%%%%%%%%%%%%%%%%%%%%%%%%%%%%%%%%%%%%%%%%%%%%%%%%%%%%%%%%%%%%%%%%%%%%%%%%%%%%%%%%%%%%%%%%%%%%%%%%%%%%%%%%%%%%%%%%%%%%%%%%%%%%%%%%%%%%%%%%%%%%%%%

\subsection{Heat capacity}

Lastly, we present the heat capacity of the system. This quantity provides insights into how the system responds to temperature changes. The heat capacity is crucial for understanding the energy absorption and release as the temperature varies
\ie
\begin{split}
\label{CCC}
C_{V} (r,r_{0}) & = -\int^{\infty}_{0} \frac{E \sqrt{-E^2} \left(\beta  E-\left(e^{\beta  E}-1\right) \ln \left(1-e^{-\beta  E}\right)\right)}{\left(e^{\beta  E}-1\right) \left(\frac{r-r_{0} \left(\frac{r_{0}}{r}\right)^{\frac{3 \lambda +(5 \lambda +3) w+1}{-\lambda +(\lambda +3) w+1}}}{(\lambda +1) r}\right)^{3/2}} \\
& = \frac{4 \pi ^4 T^3}{15 \left(\frac{r_{0} \left(\frac{r_{0}}{r}\right)^{\frac{3 \lambda +(5 \lambda +3) w+1}{-\lambda +(\lambda +3) w+1}}-r}{(\lambda +1) r}\right)^{3/2}}.
\end{split}
\fe

\begin{figure}
    \centering
     \includegraphics[scale=0.5]{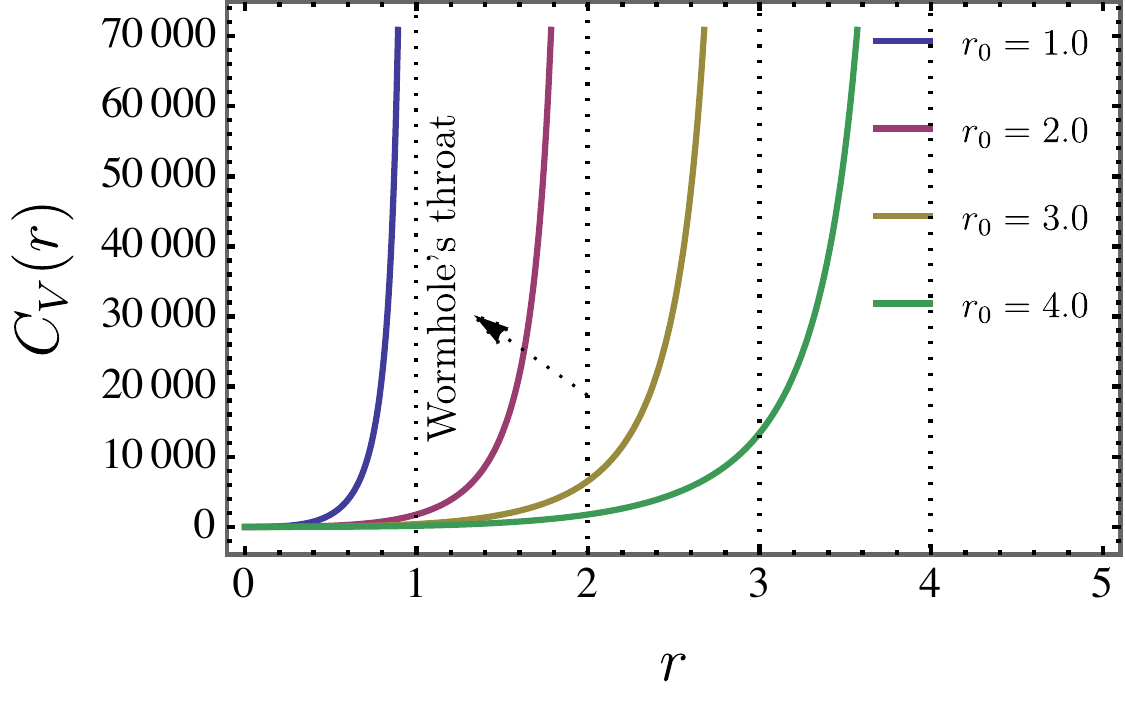}
    \caption{The heat capacity $C_{V}(r)$ as a function of $r$ for different configurations of $r_{0}$ and $T=10 K$.}
    \label{cr}
\end{figure}

The behavior of this equation is explored in Fig. \ref{cr}. Like the pressure and entropy, the heat capacity exhibits a divergence at \(r_{0}\). To further understand this, we will consider three different scenarios to analyze the heat capacity, as outlined below.

%%%%%%%%%%%%%%%%%%%%%%%%%%%%%%%%%%%%%%%%%%%%%%%%%%%%%%%%%%%%%%%%%%%%%%%%%%%%%%%%%%%%%%%%%%%%%%%%%%%%%%%%%%%%%%%%%%%%%%%%%%%%%%%%%%%%%%%%%%%%%

\subsubsection{Asymptotically far}

\begin{figure}
    \centering
     \includegraphics[scale=0.5]{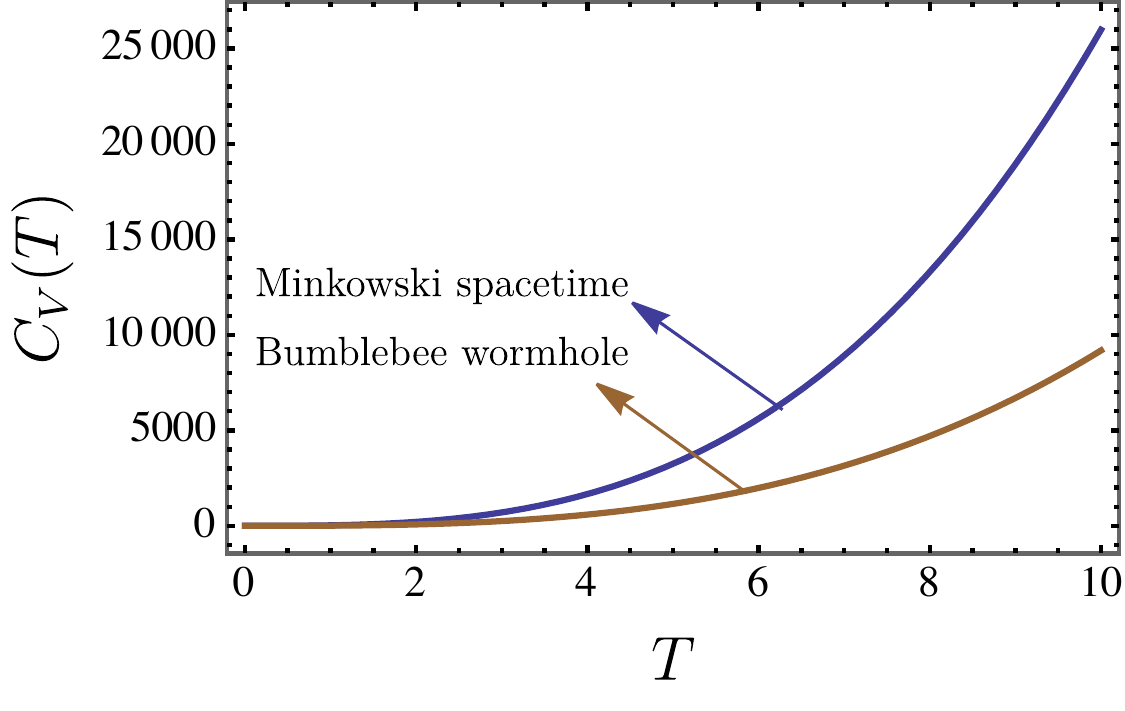}
    \caption{The comparison of the heat capacity of photon--like particles in Minkowski and bumblebee wormhole spacetimes.}
    \label{MBC}
\end{figure}

We now evaluate the limit \(\lim\limits_{r \to \infty} C_{V}(r, r_{0})\), which yields the following result:
\ie
\lim\limits_{r \to \infty}C_{V}(r,r_{0}) = C_{V} (T) = \frac{1}{15} \sqrt{2} \pi ^4 T^3.
\fe
It is essential to emphasize that this result shows a slight deviation from the one obtained for photons in Minkowski spacetime, where \(C_{V}(T) = \frac{4 \pi^4 T^3}{15}\). This difference is partly due to the fact that the bumblebee wormhole exists in a non--asymptotically flat spacetime \cite{ovgun2019exact}. To provide a clearer comparison between our results and those from Minkowski spacetime, we present Fig. \ref{MBC}.

%%%%%%%%%%%%%%%%%%%%%%%%%%%%%%%%%%%%%%%%%%%%%%%%%%%%%%%%%%%%%%%%%%%%%%%%%%%%%%%%%%%%%%%%%%%%%%%%%%%%%%%%%%%%%%%%%%%%%%%%%%%%%%%%%%%%%%%%%%%%%

\subsubsection{Close to the throat}

\begin{figure}
    \centering
     \includegraphics[scale=0.5]{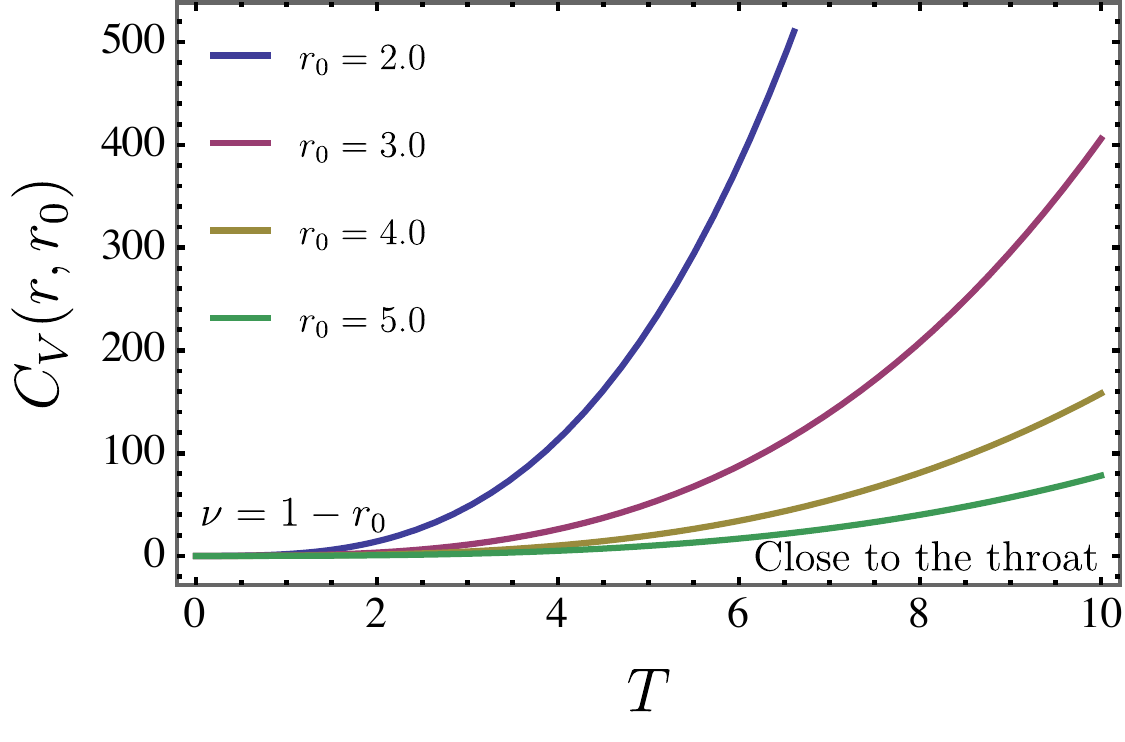}
    \caption{The heat capacity $C_{V}(r,r_{0})$ as a function of temperature $T$ for different values of the wormhole`s throat $r_{0}$, when we consider a configuration located close to it.}
    \label{closeC}
\end{figure}

In this section, we focus on the thermodynamic behavior of the system in the vicinity of the throat. To explore this, we redefine the radial coordinate as \(r = r_{0} - \nu\), where \(\nu\) represents a small deviation from the throat. Specifically, we choose \(\nu = 1 - r_{0}\) for this analysis. Under these assumptions, the heat capacity is given by the following expression:
\ie
C_{V} (r,r_{0}) = \frac{4 \pi ^4 T^3}{15 \left(\frac{-\nu +r_{0} \left(\frac{r_{0}}{\nu +r_{0}}\right)^{\frac{3 \lambda +(5 \lambda +3) w+1}{-\lambda +(\lambda +3) w+1}}-r_{0}}{(\lambda +1) (\nu +r_{0})}\right)^{3/2}}.
\fe
We now analyze the behavior of this quantity in Fig. \ref{closeC}. The plots corresponding to various values of \(r_{0}\). However, as we will explore in the next subsection, this pattern is markedly different from the behavior observed in configurations located much closer to the wormhole's throat.

%%%%%%%%%%%%%%%%%%%%%%%%%%%%%%%%%%%%%%%%%%%%%%%%%%%%%%%%%%%%%%%%%%%%%%%%%%%%%%%%%%%%%%%%%%%%%%%%%%%%%%%%%%%%%%%%%%%%%%%%%%%%%%%%%%%%%%%%%%%%%

\subsubsection{Very close to the throat}

\begin{figure}
    \centering
     \includegraphics[scale=0.425]{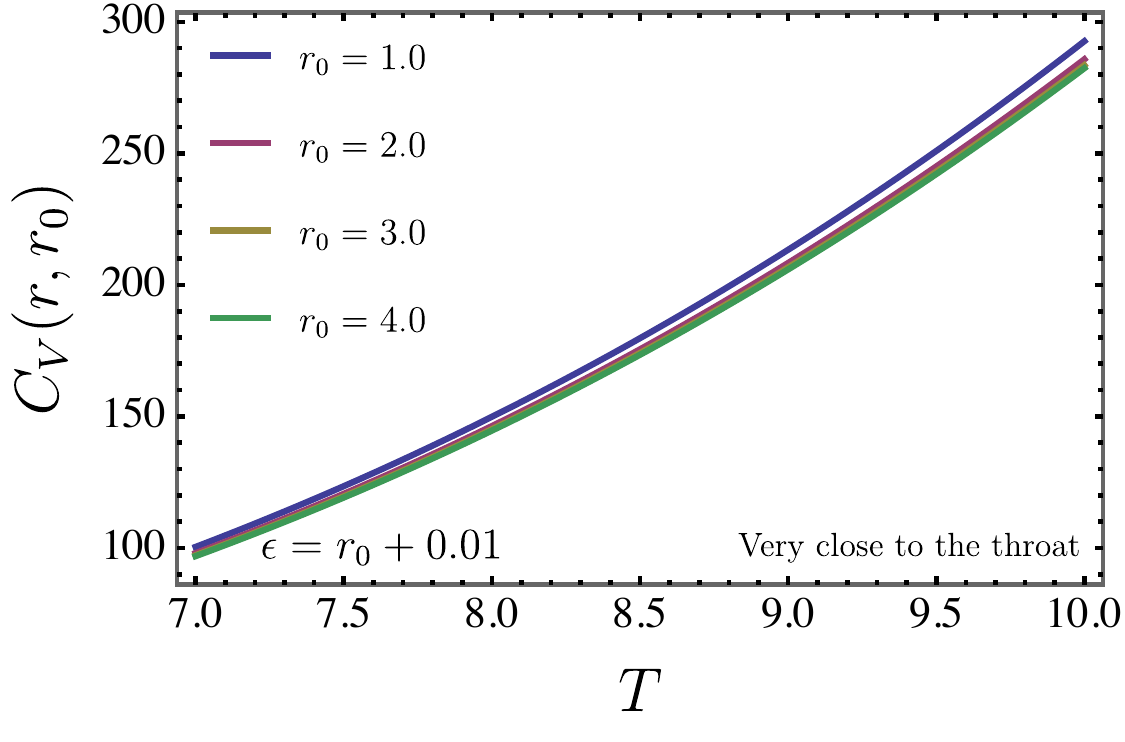}
     \includegraphics[scale=0.42]{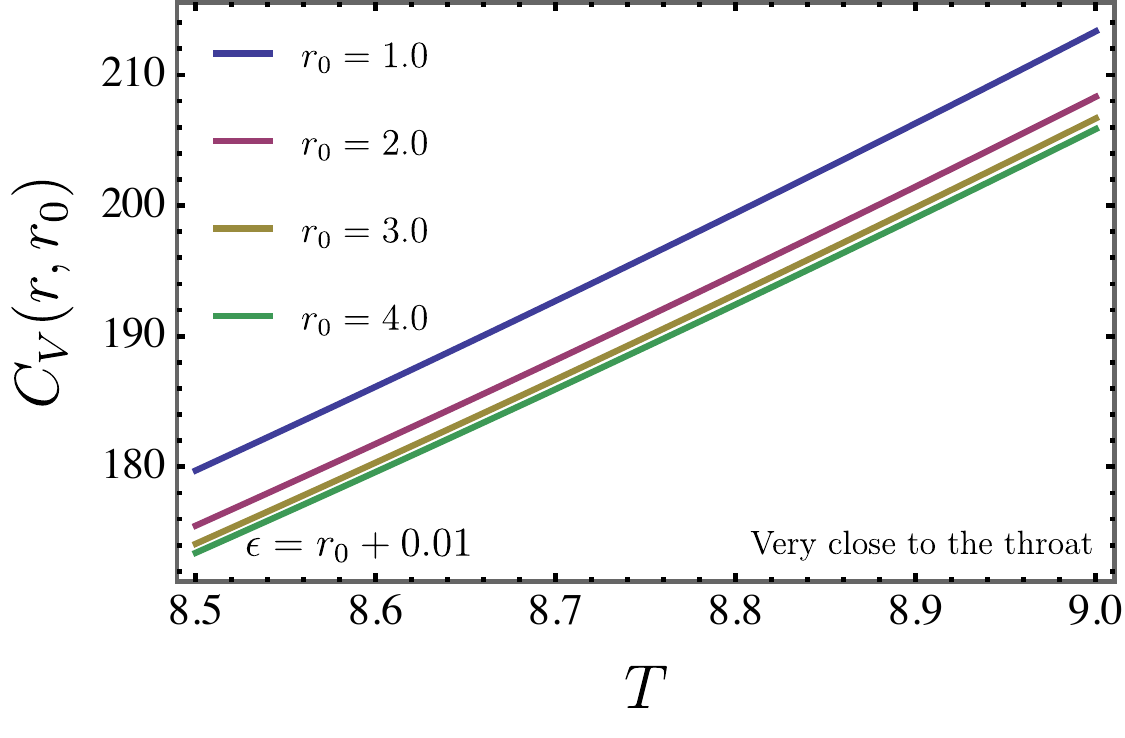}
    \caption{The heat capacity $C_{V}(r,r_{0})$ as a function of temperature $T$ for different values of the wormhole`s throat $r_{0}$, when we consider a configuration located very close to it.}
    \label{verycloseC}
\end{figure}

In this section, we examine the behavior near the throat of the bumblebee wormhole. A pertinent question might be: why not perform the calculations exactly at the throat? The answer lies in the fact that such calculations lead to divergences in this case. As a result, we choose to carry out the analysis near \(r_{0}\). To do so, we define \(r = \epsilon + r_{0}\), where \(\epsilon\) is a small parameter. Expanding Eq. (\ref{CCC}) under these conditions gives:
\ie
C_{V} (r,r_{0}) = \frac{1}{256} \pi ^4 \left(\frac{1}{r_{0}}\right)^{3/2} T^3 \epsilon ^{3/2}+\frac{\pi ^4 T^3}{30 \left(\frac{1}{r_{0}}\right)^{3/2} \epsilon ^{3/2}}+\frac{13}{320} \pi ^4 \sqrt{\frac{1}{r_{0}}} T^3 \sqrt{\epsilon }-\frac{3 \left(\pi ^4 T^3\right)}{40 \sqrt{\frac{1}{r_{0}}} \sqrt{\epsilon }}.
\fe
By expanding the expression up to second order in the parameter \(\epsilon\), we derive the necessary terms. For a clearer understanding of this result, Fig. \ref{verycloseC} is provided. In this plot, the pressure is computed near \(r_{0}\) with \(\epsilon = r_{0} + 0.01\), considering several values of \(r_{0}\). Unlike the earlier analysis near the throat, this figure reveals a clear divergence for smaller \(\epsilon\), as anticipated.

\section{Conclusion}

This work was devoted to addressing various phenomena associated with the geometry of the bumblebee traversable wormhole, with a particular focus on photon--like particles in this framework. Initially, we derived the relation between the Hamiltonian and the momentum of the theory for the most general case of a spherically symmetric spacetime. This led to the emergence of a modified dispersion relation. In broad terms, we calculated the index of refraction $n(r)$, the group velocity $v_{g}$, the time delay $\Delta t(d)$, modified distances $D^{\text{modified}}_{c}$, and the interparticle potential $V(r)$. Additionally, we explored the thermodynamic properties of the system, specifically pressure $P(r,r_{0})$, mean energy $U(r,r_{0})$, entropy $S(r,r_{0})$, and heat capacity $C_{V}(r,r_{0})$, considering three regions of interest: asymptotically far, near the throat, and very close to the throat.

In particular, $n(r)$ exhibited divergences at the throat $r_{0}$. For increasing values of $r_{0}$, the magnitude of $n(r)$ increased, while as $r$ progressed, $n(r)$ diminished. Conversely, the group velocity increased as $r$ advanced. At asymptotically far distances ($r \to \infty$), $v_{g}$ approached a constant value of $\sqrt{2}$. Divergences in $v_{g}$ were also observed at the wormhole's throat. To support the time delay calculations, we applied the results to an astrophysical scenario. For the interparticle potential, we employed the Green’s function method to obtain the results. For massive particles, $V(r)$ resulted in a combination of Yukawa-- and Coulomb--like interactions, whereas for photon--like particles, $V_{0}(r)$ displayed a Coulomb--like interaction. This indicated that photon--photon interactions were naturally possible within the bumblebee wormhole geometry. Moreover, at very far distances ($r \to \infty$), $V_{0}(r)$ approached a constant value of $1/8\pi$. Both $V(r)$ and $V_{0}(r)$ showed divergences at the wormhole's throat.

In addition to these analyses, the Hawking temperature was derived using the trapping horizon method. In this context, the negative values were obtained for it, suggesting that particles emerging from the wormhole exhibit characteristics similar to those of phantom energy, which is associated with negative temperatures \cite{Martin-Moruno:2009rpi,Gonzalez-Diaz:2004clm,Ovgun:2015taa,Rehman:2020myc}.

Considering the analysis through ensemble theory, the thermodynamic properties also depended on the parameters governing the bumblebee wormhole. Notably, all quantities --- $P(r,r_{0})$, $U(r,r_{0})$, $S(r,r_{0})$, and $C_{V}(r,r_{0})$ --- were derived analytically. Divergences appeared in all these quantities at the throat as well, and they were examined in the three regions: far from the throat, near the throat, and very close to the throat. Additionally, our findings were juxtaposed with the thermal characteristics of massless particles in Minkowski spacetime. %{\color{red} Crucially, the incorporation of interactions effectively mitigated the divergences linked to the wormhole geometry, especially close to the throat and within the low-temperature regime.

%The comparative evaluation highlights the indispensable role of the interaction term in yielding thermodynamic predictions that are physically meaningful. Its incorporation stabilizes vital thermodynamic properties, mitigates divergences in proximity to the wormhole throat, and introduces long-range corrections that are essential for precise modeling. Without the interaction term, the model is constrained to a localized description primarily influenced by thermal effects, thereby neglecting persistent contributions in the asymptotic regime. Consequently, the inclusion of the interaction term $U\left( V,\bar{n}\right)$ is not merely a theoretical enhancement but a necessary element to maintain the stability and precision of the bumblebee wormhole model. This methodology furnishes a comprehensive and robust representation of particle dynamics within such an exotic spacetime framework. Even though we consider only first-order approximations, incorporating higher-order terms may be necessary to further mitigate divergences and enhance model precision. While the current analysis employs solely first-order approximations, the implementation of higher-order terms may be imperative to further attenuate divergences and improve the precision of the model.}

From a broader perspective, it is worth considering the analysis of other spherically symmetric configurations in the context of Lorentz violation. Notable examples include the bumblebee wormhole solution within the framework of metric--affine formalism, as well as the black hole solution in Kalb--Ramond gravity.

%%%%%%%%%%%%%%%%%%%%%%%%%%%%%%%%%%%%%%%%%%%%%%%%%%%%%%%%%%%%%%%%%%%%%%%%%%%%%%%%%%%%%%%%%%%%%%%%%%%%%%%%%%%%%%%%%%%%%%%%%%%%%%%%%%%%%%%%%%%%%%%%%%%%%%%%%%%%%%%%%%%%%%%%%%%%%%%%%%%%%%%%%%%%%%%%%%%%%%%%%%%%%%%%%%%%%%%%%%%%%%%%%%%%%%%%%%%%%%%%%%%%%%%%%%%%%%%%%%%%%%%%

\section*{Acknowledgments}
\hspace{0.5cm}

A. A. Araújo Filho is supported by Conselho Nacional de Desenvolvimento Cient\'{\i}fico e Tecnol\'{o}gico (CNPq) and Fundação de Apoio à Pesquisa do Estado da Paraíba (FAPESQ) -- [150891/2023-7]. The authors also are indebted with Professor J. Furtado for the help with the wormhole illustration. A. {\"O}. would like to acknowledge the contribution of the COST Action CA21106 - COSMIC WISPers in the Dark Universe: Theory, astrophysics and experiments (CosmicWISPers) and the COST Action CA22113 - Fundamental challenges in theoretical physics (THEORY-CHALLENGES). We also thank TUBITAK and SCOAP3 for their support.

%%%%%%%%%%%%%%%%%%%%%%%%%%%%%%%%%%%%%%%%%%%%%%%%%%%%%%%%%%%%%%%%%%%%%%%%%%%%%%%%%%%%%%%%%%
\section{Data Availability Statement}

Data Availability Statement: No Data associated in the manuscript

%%%%%%%%%%%%%%%%%%%%%%%%%%%%%%%%%%%%%%%%%%%%%%%%%%%%%%%%%%%%%%%%%%%%%%%%%%%%%%%%%%%%%%%%%%

\bibliographystyle{ieeetr}
\bibliography{main}

\begin{thebibliography}{100}

\bibitem{flamm1916beitrage}
L.~Flamm, {\em Beitr{\"a}ge zur Einsteinschen gravitationstheorie}.
\newblock Hirzel, 1916.

\bibitem{einstein1935particle}
A.~Einstein and N.~Rosen, ``The particle problem in the general theory of
  relativity,'' {\em Physical Review}, vol.~48, no.~1, p.~73, 1935.

\bibitem{wheeler1955geons}
J.~A. Wheeler, ``Geons,'' {\em Physical Review}, vol.~97, no.~2, p.~511, 1955.

\bibitem{ovgun2019exact}
A.~\"Ovg\"un, K.~Jusufi, and I.~Sakall\i{}, ``{Exact traversable wormhole
  solution in bumblebee gravity},'' {\em Phys. Rev. D}, vol.~99, no.~2,
  p.~024042, 2019.

\bibitem{Ovgun:2018fnk}
A.~\"Ovg\"un, ``{Light deflection by Damour-Solodukhin wormholes and
  Gauss-Bonnet theorem},'' {\em Phys. Rev. D}, vol.~98, no.~4, p.~044033, 2018.

\bibitem{Javed:2019qyg}
W.~Javed, R.~Babar, and A.~\"Ovg\"un, ``{The effect of the Brane-Dicke coupling
  parameter on weak gravitational lensing by wormholes and naked
  singularities},'' {\em Phys. Rev. D}, vol.~99, no.~8, p.~084012, 2019.

\bibitem{Ovgun:2020yuv}
A.~\"Ovg\"un, ``{Weak Deflection Angle of Black-bounce Traversable Wormholes
  Using Gauss-Bonnet Theorem in the Dark Matter Medium},'' {\em Turk. J.
  Phys.}, vol.~44, no.~5, pp.~465--471, 2020.

\bibitem{Panyasiripan:2024kyu}
P.~Panyasiripan, N.~Kaewkhao, P.~Channuie, and A.~\"Ovg\"un, ``{Traversable
  wormholes in minimally geometrical deformed trace-free gravity using
  gravitational decoupling},'' {\em Nucl. Phys. B}, vol.~1004, p.~116563, 2024.

\bibitem{visser1995lorentzian}
M.~Visser, ``Lorentzian wormholes. from einstein to hawking,'' {\em Woodbury},
  1995.

\bibitem{bouhmadi2014wormholes}
M.~Bouhmadi-Lopez, F.~S. Lobo, and P.~Martin-Moruno, ``Wormholes minimally
  violating the null energy condition,'' {\em Journal of Cosmology and
  Astroparticle Physics}, vol.~2014, no.~11, p.~007, 2014.

\bibitem{cataldo2009evolving}
M.~Cataldo, S.~Del~Campo, P.~Minning, and P.~Salgado, ``Evolving lorentzian
  wormholes supported by phantom matter and cosmological constant,'' {\em
  Physical Review D}, vol.~79, no.~2, p.~024005, 2009.

\bibitem{b1}
M.~Churilova, R.~Konoplya, and A.~Zhidenko, ``Arbitrarily long-lived
  quasinormal modes in a wormhole background,'' {\em Physics Letters B},
  vol.~802, p.~135207, 2020.

\bibitem{b3}
R.~Konoplya, ``How to tell the shape of a wormhole by its quasinormal modes,''
  {\em Physics Letters B}, vol.~784, pp.~43--49, 2018.

\bibitem{verlindearaujo}
A.~Ara{\'u}jo~Filho, ``Analysis of a regular black hole in verlinde’s
  gravity,'' {\em Classical and Quantum Gravity}, vol.~41, no.~1, p.~015003,
  2023.

\bibitem{STR7}
G.~Rubtsov, P.~Satunin, and S.~Sibiryakov, ``The influence of lorentz violation
  on uhe photon detection,'' {\em CPT and Lorentz Symmetry}, p.~192–195,
  2014.

\bibitem{STR2}
H.~P. Robertson, ``Postulate versus observation in the special theory of
  relativity,'' {\em Rev. Mod. Phys.}, vol.~21, pp.~378--382, 1949.

\bibitem{STR1}
S.~Judes and M.~Visser, ``Conservation laws in ``doubly special relativity'',''
  {\em Phys. Rev. D}, vol.~68, p.~045001, 2003.

\bibitem{amarilo2024gravitational}
K.~M. Amarilo, M.~B. Ferreira~Filho, A.~A. Ara{\'u}jo~Filho, and J.~A. A.~S.
  Reis, ``Gravitational waves effects in a lorentz--violating scenario,'' {\em
  Physics Letters B}, p.~138785, 2024.

\bibitem{STR5}
C.~M. Reyes, L.~F. Urrutia, and J.~D. Vergara, ``Quantization of the
  myers-pospelov model: The photon sector interacting with standard fermions as
  a perturbation of qed,'' {\em Phys. Rev. D}, vol.~78, p.~125011, 2008.

\bibitem{hees2016}
A.~Hees, Q.~G. Bailey, A.~Bourgoin, P.-L. Bars, C.~Guerlin, L.~Poncin-Lafitte,
  {\em et~al.}, ``Tests of lorentz symmetry in the gravitational sector,'' {\em
  Universe}, vol.~2, no.~4, p.~30, 2016.

\bibitem{liberati2013}
S.~Liberati, ``Tests of lorentz invariance: a 2013 update,'' {\em Class. Quant.
  Grav.}, vol.~30, no.~13, p.~133001, 2013.

\bibitem{tasson2014}
J.~D. Tasson, ``What do we know about lorentz invariance?,'' {\em Rep. Progr.
  Phys.}, vol.~77, no.~6, p.~062901, 2014.

\bibitem{rovelli2004}
C.~Rovelli, {\em Quantum gravity}.
\newblock Cambridge university press, 2004.

\bibitem{New2}
V.~A. Kosteleck\'y and S.~Samuel, ``Phenomenological gravitational constraints
  on strings and higher-dimensional theories,'' {\em Phys. Rev. Lett.},
  vol.~63, pp.~224--227, 1989.

\bibitem{New1}
V.~A. Kosteleck\'y and S.~Samuel, ``Spontaneous breaking of lorentz symmetry in
  string theory,'' {\em Phys. Rev. D}, vol.~39, pp.~683--685, 1989.

\bibitem{New7}
M.~Bojowald, H.~A. Morales-T\'ecotl, and H.~Sahlmann, ``Loop quantum gravity
  phenomenology and the issue of lorentz invariance,'' {\em Phys. Rev. D},
  vol.~71, p.~084012, 2005.

\bibitem{New6}
R.~Gambini and J.~Pullin, ``Nonstandard optics from quantum space-time,'' {\em
  Phys. Rev. D}, vol.~59, p.~124021, 1999.

\bibitem{New9}
S.~M. Carroll, J.~A. Harvey, V.~A. Kosteleck\'y, C.~D. Lane, and T.~Okamoto,
  ``Noncommutative field theory and lorentz violation,'' {\em Phys. Rev.
  Lett.}, vol.~87, p.~141601, 2001.

\bibitem{New8}
G.~Amelino-Camelia and S.~Majid, ``Waves on noncommutative space--time and
  gamma-ray bursts,'' {\em International Journal of Modern Physics A}, vol.~15,
  no.~27, pp.~4301--4323, 2000.

\bibitem{Modesto:2011kw}
L.~Modesto, ``{Super-renormalizable Quantum Gravity},'' {\em Phys. Rev. D},
  vol.~86, p.~044005, 2012.

\bibitem{New11}
S.~Bernadotte and F.~R. Klinkhamer, ``Bounds on length scales of classical
  spacetime foam models,'' {\em Phys. Rev. D}, vol.~75, p.~024028, 2007.

\bibitem{New10}
F.~R. Klinkhamer and C.~Rupp, ``Spacetime foam, cpt anomaly, and photon
  propagation,'' {\em Phys. Rev. D}, vol.~70, p.~045020, 2004.

\bibitem{New15}
K.~Ghosh and F.~Klinkhamer, ``Anomalous lorentz and cpt violation from a local
  chern–simons-like term in the effective gauge-field action,'' {\em Nucl.
  Phys. B}, vol.~926, pp.~335 -- 369, 2018.

\bibitem{New14}
F.~Klinkhamer and J.~Schimmel, ``Cpt anomaly: a rigorous result in four
  dimensions,'' {\em Nucl. Phys. B}, vol.~639, no.~1, pp.~241 -- 262, 2002.

\bibitem{New16}
P.~Ho\ifmmode~\check{r}\else \v{r}\fi{}ava, ``Quantum gravity at a lifshitz
  point,'' {\em Phys. Rev. D}, vol.~79, p.~084008, 2009.

\bibitem{sv2}
A.~Casalino, M.~Rinaldi, L.~Sebastiani, and S.~Vagnozzi, ``Alive and well:
  mimetic gravity and a higher-order extension in light of gw170817,'' {\em
  Class. Quant. Grav.}, vol.~36, no.~1, p.~017001, 2018.

\bibitem{sv1}
G.~Cognola, R.~Myrzakulov, L.~Sebastiani, S.~Vagnozzi, and S.~Zerbini,
  ``Covariant ho{\v{r}}ava-like and mimetic horndeski gravity: cosmological
  solutions and perturbations,'' {\em Class. Quant. Grav.}, vol.~33, no.~22,
  p.~225014, 2016.

\bibitem{Kluson:2011rs}
J.~Kluson, S.~Nojiri, and S.~D. Odintsov, ``{Covariant Lagrange multiplier
  constrained higher derivative gravity with scalar projectors},'' {\em Phys.
  Lett. B}, vol.~701, pp.~117--126, 2011.

\bibitem{Kluson:2010za}
J.~Kluson, S.~Nojiri, S.~D. Odintsov, and D.~Saez-Gomez, ``{$U$(1) Invariant
  $F(\tilde{R})$ Horava-Lifshitz Gravity},'' {\em Eur. Phys. J. C}, vol.~71,
  p.~1690, 2011.

\bibitem{Nojiri:2010kx}
S.~Nojiri and S.~D. Odintsov, ``{Covariant power-counting renormalizable
  gravity: Lorentz symmetry breaking and accelerating early-time FRW
  universe},'' {\em Phys. Rev. D}, vol.~83, p.~023001, 2011.

\bibitem{Nojiri:2010tv}
S.~Nojiri and S.~D. Odintsov, ``{A proposal for covariant renormalizable field
  theory of gravity},'' {\em Phys. Lett. B}, vol.~691, pp.~60--64, 2010.

\bibitem{Carloni:2010nx}
S.~Carloni, M.~Chaichian, S.~Nojiri, S.~D. Odintsov, M.~Oksanen, and
  A.~Tureanu, ``{Modified first-order Horava-Lifshitz gravity: Hamiltonian
  analysis of the general theory and accelerating FRW cosmology in power-law
  F(R) model},'' {\em Phys. Rev. D}, vol.~82, p.~065020, 2010.
\newblock [Erratum: Phys.Rev.D 85, 129904 (2012)].

\bibitem{CFJ}
S.~M. Carroll, G.~B. Field, and R.~Jackiw, ``Limits on a lorentz-and
  parity-violating modification of electrodynamics,'' {\em Phys. Rev. D},
  vol.~41, no.~4, p.~1231, 1990.

\bibitem{aether}
S.~M. Carroll and H.~Tam, ``Aether compactification,'' {\em Phys. Rev. D},
  vol.~78, no.~4, p.~044047, 2008.

\bibitem{colladay1998lorentz}
D.~Colladay and V.~A. Kosteleck{\`y}, ``Lorentz-violating extension of the
  standard model,'' {\em Phys. Rev. D}, vol.~58, no.~11, p.~116002, 1998.

\bibitem{bluhm2021gravity}
R.~Bluhm and Y.~Yang, ``Gravity with explicit diffeomorphism breaking,'' {\em
  Symmetry}, vol.~13, no.~4, p.~660, 2021.

\bibitem{Filho:2022yrk}
A.~A. Araújo~Filho, J.~R. Nascimento, A.~Y. Petrov, and P.~J. Porf\'\i{}rio,
  ``{Vacuum solution within a metric-affine bumblebee gravity},'' {\em Phys.
  Rev. D}, vol.~108, no.~8, p.~085010, 2023.

\bibitem{kostelecky2004gravity}
V.~A. Kosteleck{\`y}, ``Gravity, lorentz violation, and the standard model,''
  {\em Phys. Rev. D}, vol.~69, no.~10, p.~105009, 2004.

\bibitem{araujo2024gravitational}
A.~A. Ara{\'u}jo~Filho, H.~Hassanabadi, N.~Heidari, J.~Kriz, and S.~Zare,
  ``Gravitational traces of bumblebee gravity in metric--affine formalism,''
  {\em Classical and Quantum Gravity}, vol.~41, no.~5, p.~055003, 2024.

\bibitem{kostelecky2021backgrounds}
V.~A. Kosteleck{\`y} and Z.~Li, ``Backgrounds in gravitational effective field
  theory,'' {\em Phys. Rev. D}, vol.~103, no.~2, p.~024059, 2021.

\bibitem{araujo2024exact}
A.~A. Ara{\'u}jo~Filho, J.~R. Nascimento, A.~Y. Petrov, P.~J. Porf{\i}rio, {\em
  et~al.}, ``An exact stationary axisymmetric vacuum solution within a
  metric--affine bumblebee gravity,'' {\em Journal of Cosmology and
  Astroparticle Physics}, vol.~07, p.~004, 2024.

\bibitem{bluhm2005spontaneous}
R.~Bluhm and V.~A. Kosteleck{\`y}, ``Spontaneous lorentz violation,
  nambu-goldstone modes, and gravity,'' {\em Physical Review D—Particles,
  Fields, Gravitation, and Cosmology}, vol.~71, no.~6, p.~065008, 2005.

\bibitem{araujo2024exploring}
A.~A. Ara{\'u}jo~Filho, J.~A. A.~S. Reis, and H.~Hassanabadi, ``Exploring
  antisymmetric tensor effects on black hole shadows and quasinormal
  frequencies,'' {\em Journal of Cosmology and Astroparticle Physics},
  vol.~2024, no.~05, p.~029, 2024.

\bibitem{bluhm2023spontaneous}
R.~Bluhm and Y.~Zhi, ``Spontaneous and explicit spacetime symmetry breaking in
  einstein--cartan theory with background fields,'' {\em Symmetry}, vol.~16,
  no.~1, p.~25, 2023.

\bibitem{bluhm2008spontaneous}
R.~Bluhm, S.-H. Fung, and V.~A. Kosteleck{\`y}, ``Spontaneous lorentz and
  diffeomorphism violation, massive modes, and gravity,'' {\em Physical Review
  D—Particles, Fields, Gravitation, and Cosmology}, vol.~77, no.~6,
  p.~065020, 2008.

\bibitem{pp7}
M.~D. Seifert, ``Vector models of gravitational lorentz symmetry breaking,''
  {\em Physical Review D—Particles, Fields, Gravitation, and Cosmology},
  vol.~79, no.~12, p.~124012, 2009.

\bibitem{pp2}
K.~O’Neal-Ault, Q.~G. Bailey, and N.~A. Nilsson, ``3+ 1 formulation of the
  standard model extension gravity sector,'' {\em Physical Review D}, vol.~103,
  no.~4, p.~044010, 2021.

\bibitem{pp4}
R.~Bluhm, H.~Bossi, and Y.~Wen, ``Gravity with explicit spacetime symmetry
  breaking and the standard model extension,'' {\em Physical Review D},
  vol.~100, no.~8, p.~084022, 2019.

\bibitem{araujo2022thermal}
A.~A. Ara{\'u}jo~Filho, {\em Thermal aspects of field theories}.
\newblock Amazon. com, 2022.

\bibitem{amelino2001testable}
G.~Amelino-Camelia, ``Testable scenario for relativity with minimum length,''
  {\em Physics Letters B}, vol.~510, no.~1-4, pp.~255--263, 2001.

\bibitem{kostelecky2011data}
V.~A. Kosteleck{\`y} and N.~Russell, ``Data tables for lorentz and c p t
  violation,'' {\em Rev. Mod. Phys.}, vol.~83, no.~1, p.~11, 2011.

\bibitem{lsb1}
M.~Visser, ``Lorentz symmetry breaking as a quantum field theory regulator,''
  {\em Physical Review D—Particles, Fields, Gravitation, and Cosmology},
  vol.~80, no.~2, p.~025011, 2009.

\bibitem{lsb2}
R.~Bluhm, ``Breaking lorentz symmetry,'' {\em Physics World}, vol.~17, no.~3,
  p.~41, 2004.

\bibitem{lsb3}
B.~Knorr, ``Lorentz symmetry is relevant,'' {\em Physics Letters B}, vol.~792,
  pp.~142--148, 2019.

\bibitem{lsb4}
R.~Lehnert, ``Cpt-and lorentz-symmetry breaking: a review,'' {\em arXiv
  preprint hep-ph/0611177}, 2006.

\bibitem{lsb5}
C.~A. Hernaski, ``Spontaneous breaking of lorentz symmetry with an
  antisymmetric tensor,'' {\em Physical Review D}, vol.~94, no.~10, p.~105004,
  2016.

\bibitem{lsb6}
G.~Cuomo and S.~Zhang, ``Spontaneous symmetry breaking on surface defects,''
  {\em Journal of High Energy Physics}, vol.~2024, no.~3, pp.~1--50, 2024.

\bibitem{colladay2004statistical}
D.~Colladay and P.~McDonald, ``Statistical mechanics and lorentz violation,''
  {\em Physical Review D}, vol.~70, no.~12, p.~125007, 2004.

\bibitem{aa2021lorentz}
A.~A. Ara{\'u}jo~Filho, ``Lorentz-violating scenarios in a thermal reservoir,''
  {\em The Eur. Phys. J. Plus}, vol.~136, no.~4, pp.~1--14, 2021.

\bibitem{anacleto2018lorentz}
M.~Anacleto, F.~Brito, E.~Maciel, A.~Mohammadi, E.~Passos, W.~Santos, and
  J.~Santos, ``Lorentz-violating dimension-five operator contribution to the
  black body radiation,'' {\em Physics Letters B}, vol.~785, pp.~191--196,
  2018.

\bibitem{araujo2021thermodynamic}
A.~A. Ara{\'u}jo~Filho and R.~V. Maluf, ``Thermodynamic properties in
  higher-derivative electrodynamics,'' {\em Braz. J. Phys.}, vol.~51, no.~3,
  pp.~820--830, 2021.

\bibitem{casana2009finite}
R.~Casana, M.~M. Ferreira~Jr, J.~S. Rodrigues, and M.~R. Silva, ``Finite
  temperature behavior of the c p t-even and parity-even electrodynamics of the
  standard model extension,'' {\em Physical Review D}, vol.~80, no.~8,
  p.~085026, 2009.

\bibitem{casana2008lorentz}
R.~Casana, M.~M. Ferreira~Jr, and J.~S. Rodrigues, ``Lorentz-violating
  contributions of the carroll-field-jackiw model to the cmb anisotropy,'' {\em
  Physical Review D}, vol.~78, no.~12, p.~125013, 2008.

\bibitem{araujo2021higher}
A.~A. Ara{\'u}jo~Filho and A.~Y. Petrov, ``Higher-derivative lorentz-breaking
  dispersion relations: a thermal description,'' {\em The European Physical
  Journal C}, vol.~81, no.~9, p.~843, 2021.

\bibitem{aguirre2021lorentz}
A.~Aguirre, G.~Flores-Hidalgo, R.~Rana, and E.~Souza, ``The lorentz-violating
  real scalar field at thermal equilibrium,'' {\em Eur. Phys. J. C}, vol.~81,
  no.~5, p.~459, 2021.

\bibitem{reis2021thermal}
J.~A. A.~S. Reis {\em et~al.}, ``Thermal aspects of interacting quantum gases
  in lorentz-violating scenarios,'' {\em Eur. Phys. J. Plus}, vol.~136, no.~3,
  p.~310, 2021.

\bibitem{Mariz:2011ed}
T.~Mariz, J.~R. Nascimento, and A.~Y. Petrov, ``{On the perturbative generation
  of the higher-derivative Lorentz-breaking terms},'' {\em Phys. Rev. D},
  vol.~85, p.~125003, 2012.

\bibitem{petrov2021bouncing2}
A.~A. Ara{\'u}jo~Filho and A.~Y. Petrov, ``Bouncing universe in a heat bath,''
  {\em Int. J. Mod. Phys. A}, vol.~36, no.~34 \& 35, p.~2150242, 2021.

\bibitem{aaa2021thermodynamics}
A.~A. Ara{\'u}jo~Filho, ``Thermodynamics of massless particles in curved
  spacetime,'' {\em arXiv preprint arXiv:2201.00066}, 2021.

\bibitem{casana2018exact}
R.~Casana, A.~Cavalcante, F.~Poulis, and E.~Santos, ``Exact schwarzschild-like
  solution in a bumblebee gravity model,'' {\em Physical Review D}, vol.~97,
  no.~10, p.~104001, 2018.

\bibitem{morris1988wormholes}
M.~S. Morris and K.~S. Thorne, ``Wormholes in spacetime and their use for
  interstellar travel: A tool for teaching general relativity,'' {\em American
  Journal of Physics}, vol.~56, no.~5, pp.~395--412, 1988.

\bibitem{nandi2016stability}
K.~K. Nandi, A.~A. Potapov, R.~Izmailov, A.~Tamang, and J.~C. Evans,
  ``Stability and instability of ellis and phantom wormholes: Are there
  ghosts?,'' {\em Physical Review D}, vol.~93, no.~10, p.~104044, 2016.

\bibitem{amelino2013quantum}
G.~Amelino-Camelia, ``Quantum-spacetime phenomenology,'' {\em Living Reviews in
  Relativity}, vol.~16, pp.~1--137, 2013.

\bibitem{ling2006modified}
Y.~Ling, B.~Hu, and X.~Li, ``Modified dispersion relations and black hole
  physics,'' {\em Physical Review D}, vol.~73, no.~8, p.~087702, 2006.

\bibitem{smolin2006case}
L.~Smolin, ``The case for background independence,'' {\em The structural
  foundations of quantum gravity}, pp.~196--239, 2006.

\bibitem{kowalski2005introduction}
J.~Kowalski-Glikman, ``Introduction to doubly special relativity,'' in {\em
  Planck Scale Effects in Astrophysics and Cosmology}, pp.~131--159, Springer,
  2005.

\bibitem{mattingly2005modern}
D.~Mattingly, ``Modern tests of lorentz invariance,'' {\em Living Reviews in
  relativity}, vol.~8, no.~1, pp.~1--84, 2005.

\bibitem{girelli2009emergence}
F.~Girelli, S.~Liberati, and L.~Sindoni, ``Emergence of lorentzian signature
  and scalar gravity,'' {\em Physical Review D}, vol.~79, no.~4, p.~044019,
  2009.

\bibitem{jacob2008lorentz}
U.~Jacob and T.~Piran, ``Lorentz-violation-induced arrival delays of
  cosmological particles,'' {\em Journal of Cosmology and Astroparticle
  Physics}, vol.~2008, no.~01, p.~031, 2008.

\bibitem{galaverni2008lorentz}
M.~Galaverni and G.~Sigl, ``Lorentz violation for photons and ultrahigh-energy
  cosmic rays,'' {\em Physical review letters}, vol.~100, no.~2, p.~021102,
  2008.

\bibitem{oliveira2019quasinormal}
R.~Oliveira, D.~Dantas, V.~Santos, and C.~Almeida, ``Quasinormal modes of
  bumblebee wormhole,'' {\em Classical and Quantum Gravity}, vol.~36, no.~10,
  p.~105013, 2019.

\bibitem{blackledge2005digital}
J.~M. Blackledge, {\em Digital image processing: mathematical and computational
  methods}.
\newblock Elsevier, 2005.

\bibitem{gradshteyn2014table}
I.~S. Gradshteyn and I.~M. Ryzhik, {\em Table of integrals, series, and
  products}.
\newblock Academic press, 2014.

\bibitem{Martin-Moruno:2009rpi}
P.~Martin-Moruno and P.~F. Gonzalez-Diaz, ``{Thermal radiation from Lorentzian
  traversable wormholes},'' {\em Phys. Rev. D}, vol.~80, p.~024007, 2009.

\bibitem{Gonzalez-Diaz:2004clm}
P.~F. Gonzalez-Diaz and C.~L. Siguenza, ``{Phantom thermodynamics},'' {\em
  Nucl. Phys. B}, vol.~697, pp.~363--386, 2004.

\bibitem{Ovgun:2015taa}
I.~Sakalli and A.~Ovgun, ``{Tunnelling of vector particles from Lorentzian
  wormholes in 3+1 dimensions},'' {\em Eur. Phys. J. Plus}, vol.~130, no.~6,
  p.~110, 2015.

\bibitem{Rehman:2020myc}
M.~Rehman and K.~Saifullah, ``{Thermodynamics of dynamical wormholes},'' {\em
  JCAP}, vol.~06, p.~020, 2021.

\bibitem{anchordoqui2003ultrahigh}
L.~Anchordoqui, T.~Paul, S.~Reucroft, and J.~Swain, ``Ultrahigh energy cosmic
  rays: The state of the art before the auger observatory,'' {\em International
  Journal of Modern Physics A}, vol.~18, no.~13, pp.~2229--2366, 2003.

\bibitem{aaa33}
J.~Furtado, H.~Hassanabadi, J.~Reis, {\em et~al.}, ``Thermal analysis of
  photon-like particles in rainbow gravity,'' {\em arXiv preprint
  arXiv:2305.08587}, 2023.

\bibitem{araujo2023thermodynamical}
A.~A. Ara{\'u}jo~Filho, J.~Furtado, J.~Reis, and J.~Silva, ``Thermodynamical
  properties of an ideal gas in a traversable wormhole,'' {\em Class. Quant.
  Grav.}, vol.~40, no.~24, p.~245001, 2023.

\bibitem{greiner2012thermodynamics}
W.~Greiner, L.~Neise, and H.~St{\"o}cker, {\em Thermodynamics and statistical
  mechanics}.
\newblock Springer Science \& Business Media, 2012.

\bibitem{isihara2013statistical}
A.~Isihara, {\em Statistical physics}.
\newblock Academic Press, 2013.

\bibitem{wannier1987statistical}
G.~H. Wannier, {\em Statistical physics}.
\newblock Courier Corporation, 1987.

\bibitem{salinas1999introduccao}
S.~R. Salinas, {\em Introdu{\c{c}}{\~a}o {\`a} f{\'\i}sica estat{\'\i}stica}.
\newblock Edusp, 1999.

\bibitem{vogt2017statistical}
J.~Vogt, ``Statistical thermodynamics,'' {\em Exam survival guide: Physical
  chemistry}, pp.~175--211, 2017.

\bibitem{mandl1991statistical}
F.~Mandl, {\em Statistical physics}, vol.~14.
\newblock John Wiley \& Sons, 1991.

\end{thebibliography}

\end{document}